# A perspective on the microscopic pressure (stress) tensor: history, current understanding, and future challenges

Kaihang Shi,[1, ‡, †] Edward Smith,[2, †] Erik E. Santiso,[1] and Keith E. Gubbins[1, †]

[1]*Department of Chemical and Biomolecular Engineering, North Carolina State University, Raleigh, North Carolina, USA*

[2]*Department of Mechanical and Aerospace Engineering, Brunel University London, Uxbridge, London, UK*

[‡]*Current address: Department of Chemical and Biological Engineering, Northwestern University, Evanston, Illinois, USA*

[†]Authors to whom the correspondence should be addressed: *kaihangshi0@gmail.com*; *edward.smith@brunel.ac.uk*; *keg@ncsu.edu*.

## Abstract

The pressure tensor (equivalent to the negative stress tensor) at both microscopic and macroscopic levels is fundamental to many aspects of engineering and science, including fluid dynamics, solid mechanics, biophysics, and thermodynamics. In this perspective paper, we review methods to calculate the microscopic pressure tensor. Connections between different pressure forms for equilibrium and non-equilibrium systems are established. We also point out several challenges in the field, including the historical controversies over the definition of the microscopic pressure tensor; the difficulties with many-body and long-range potentials; the insufficiency of software and computational tools; and the lack of experimental routes to probe the pressure tensor at the nanoscale. Possible future directions are suggested.

## Contents

# 1. Introduction

The pressure, $P$, defined as the average force $F$ per unit area acting on a surface of area $S$, is one of the primary state variables, together with the temperature, $T$, and composition, that determine the thermodynamic properties of a homogeneous system of molecules at equilibrium. In such a system the average force and the pressure are the same in all directions, $P$ is a scalar, and is well-defined even at the micro-scale. Statistical mechanics informs us that $P = P^K + P^C$, where superscript $K$ indicates the contribution due to the kinetic energy of the molecules, and $C$ indicates the configurational contribution, *i.e.* that from the intermolecular forces and any external field. For condensed phases the configurational contribution to $P$ is usually dominant, especially at low temperatures. For a perfect gas at equilibrium in the absence of an external field, or a real equilibrium gas at low enough density that the influence of intermolecular forces can be neglected, the configurational contribution is negligible, and $P = P^K = n(\mathbf{r})k_BT$, where $n(\mathbf{r})$ is the number density at position $\mathbf{r}$ and $k_B$ is the Boltzmann constant. For such a gas this relation holds even when the gas is non-uniform in density or composition, as long as the concept of a local average density, $n(\mathbf{r})$, is valid.

For more general situations, for example an inhomogeneous dense fluid, a nanoscale fluid or solid, or a non-equilibrium system, the pressure $\mathbf{P}$ is a second-order tensor, depending on the direction of both the force and of the surface it acts on. In general, $\mathbf{P}$ has 9 components $P_{\alpha\beta}$, where $P_{\alpha\beta}$ is the force per unit area in the $\beta$-direction acting on a surface element normal to the $\alpha$-direction. These components depend on position, $\mathbf{r}$, and for non-equilibrium systems they will also depend on time, $t$. Off-diagonal components are the shear pressures (shear stress) and the diagonal components are the direct pressures. In some specific types of systems, the number of non-vanishing components of $\mathbf{P}$ may be less than 9. For an inhomogeneous fluid that is at equilibrium and not under strain, for example, the off-diagonal components vanish and there are only 3 non-vanishing components. Also, the condition of mechanical (hydrostatic) equilibrium (the average rate of change of linear momentum vanishes) often provides relations between the remaining non-vanishing components.

A difficulty in many applications is that the local (microscopic) pressure tensor at some point $\mathbf{r}$ is not uniquely defined in non-equilibrium systems or in equilibrium ones that are inhomogeneous. Although the kinetic contribution, $\mathbf{P}^K$, is well-defined, as noted above, the configurational part, $\mathbf{P}^C$ is not, because the intermolecular forces themselves are non-local. Thus, while the force between molecules $i$ and $j$, located at positions $\mathbf{r}_i$ and $\mathbf{r}_j$, is well defined in general, there is no well-defined way to assign a contribution to the force acting on a surface element at some position $\mathbf{r}$. This arbitrariness in the force acting across the surface element d$S$ seems to have first been stated explicitly by Irving and Kirkwood in 1950.[1] There they stated the matter succinctly:

"*…all definitions (of the configurational pressure tensor) must have this in common – that the stress between a pair of molecules be concentrated near the line of centers. When averaging over a domain large compared with the range of intermolecular force, these differences are washed out, and the ambiguity remaining in the macroscopic stress tensor is of negligible order*."

This point was discussed in a footnote to an appendix to Irving & Kirkwood's paper, and so was not noticed widely at the time. It was of little consequence to these authors, who were primarily interested in transport processes at the macroscale. However, it is important for nanoscale systems, such as small nanoparticles, drops and fluids confined within nanoporous materials or living cells, and we expand on this later in this perspective. The non-uniqueness of the local pressure tensor, and its consequences for inhomogeneous fluids and the properties of gas-liquid interfaces, was investigated in the 1982 paper of Schofield and Henderson.[2]

While in this paper we shall mainly focus our discussion on the local pressure tensor, in fields where the primary interest is in non-equilibrium phenomena and solid mechanics, the stress tensor $\boldsymbol{\sigma}(\mathbf{r}, t)$ is usually used in place of the pressure tensor.[1] These two tensors differ only in sign:[1,3]

$$\boldsymbol{\sigma}(\mathbf{r}, t) = -\mathbf{P}(\mathbf{r}, t) \tag{1}$$

The negative sign ensures that the stress tensor definition is consistent with Newton's law of viscous flow, and that the viscosity is positive. The terms "pressure tensor" and "stress tensor" are used interchangeably in this paper and in much of the literature, the sign change being understood.

The molecular level local pressure/stress tensor has been the key to the depiction of the mechanical and thermodynamic picture of many important phenomena. Its important role is evidenced by a rapid growth of publications mentioning it (Figure 1). In biophysics, the local stress tensor has been applied to understand the mechanical properties of lipid bilayer membranes (see Figure 2a).[4–7] The structure and mechanics of the lipid membrane play a critical role in the function of proteins involved in processes of transport, signaling and mechano-transduction. The local stress tensor also enables the quantification of the mechanical state of proteins in glassy matrices.[8] Such information is pivotal to a sophisticated design and control of the lyophilization (freeze-drying) process for long-term storage and stabilization of labile biomolecules in the food and pharmaceutical industries. In material science, the stress tensor has been related to the structural deformation of the materials upon adsorption, and such a connection is useful for materials characterization.[9,10] For gas-liquid[11–13] or liquid-solid[14,15] nucleation, the pressure tensor profile provides a mechanical picture of the nucleus interfacing with the surrounding environment; such a profile is useful for calculating the Tolman length for interfacial free energy[14] and for understanding the distinct structure of the nucleus[15] (see Figure 2b). The pressure tensor profile across the interfacial region is also essential in a virial (or mechanical) route to the surface tension (see examples for planar,[3,16–19] spherical[3,11,12,20] and cylindrical[21] interfaces). For confined systems, the knowledge of the microscopic pressure tensor paves the way for understanding phase transitions in nanopores,[22–24] and for developing sophisticated equations of state for confined fluids.[25,26] Recently, the microscopic pressure tensor has provided a mechanistic understanding of high-pressure phenomena in confinement or near strongly wetting surfaces for advanced materials synthesis and enhanced chemical processing.[27] The high-pressure phenomena include enhanced chemical reactions in pores that normally require a high pressure in the bulk,[28,29] and the formation and stabilization of high-pressure phases in nanopores.[30–32] For simple non-reacting adsorption systems, high (tangential) pressures that are about three to four orders of magnitude larger than the bulk pressure were found in the adsorbed layers on carbon surfaces (see Figure 2c).[33–35] This compression effect is caused by strong attractive forces exerted on the adsorbate molecules by the surface, which leads to a tightly packed adsorbed layer near the surface, and strong repulsions between adsorbate molecules.[36–39] For more complex systems, the mechanism behind the induced high pressure in confinement is under active investigation.

The stress tensor also underpins fluid dynamics. It is the driving force in the continuum equations,[40] is necessary for the understanding of non-equilibrium molecular dynamics,[41,42] and is essential for calculating the viscosity in Newtonian fluids using both equilibrium[43,44] and non-equilibrium methods[45]. Pressure/stress can provide insight into tribology,[46] reveal the slip length[47,48] and capture many aspects of multi-phase flows, including the moving contact line for the wetting behavior (see Figure 2d)[49,50] and dynamics of active liquid interfaces.[51] These molecular details can be included in continuum-based engineering simulation, such as computational fluid dynamics (CFD), through coupled simulation[52,53] where stress coupling is useful in both fluid simulation[54] and for solid mechanics[55]. Recently, machine learning models have focused on stresses, as the central property that should be predicted for predicting atomic stress at grain boundaries[56] or in coupling to continuum models[57].

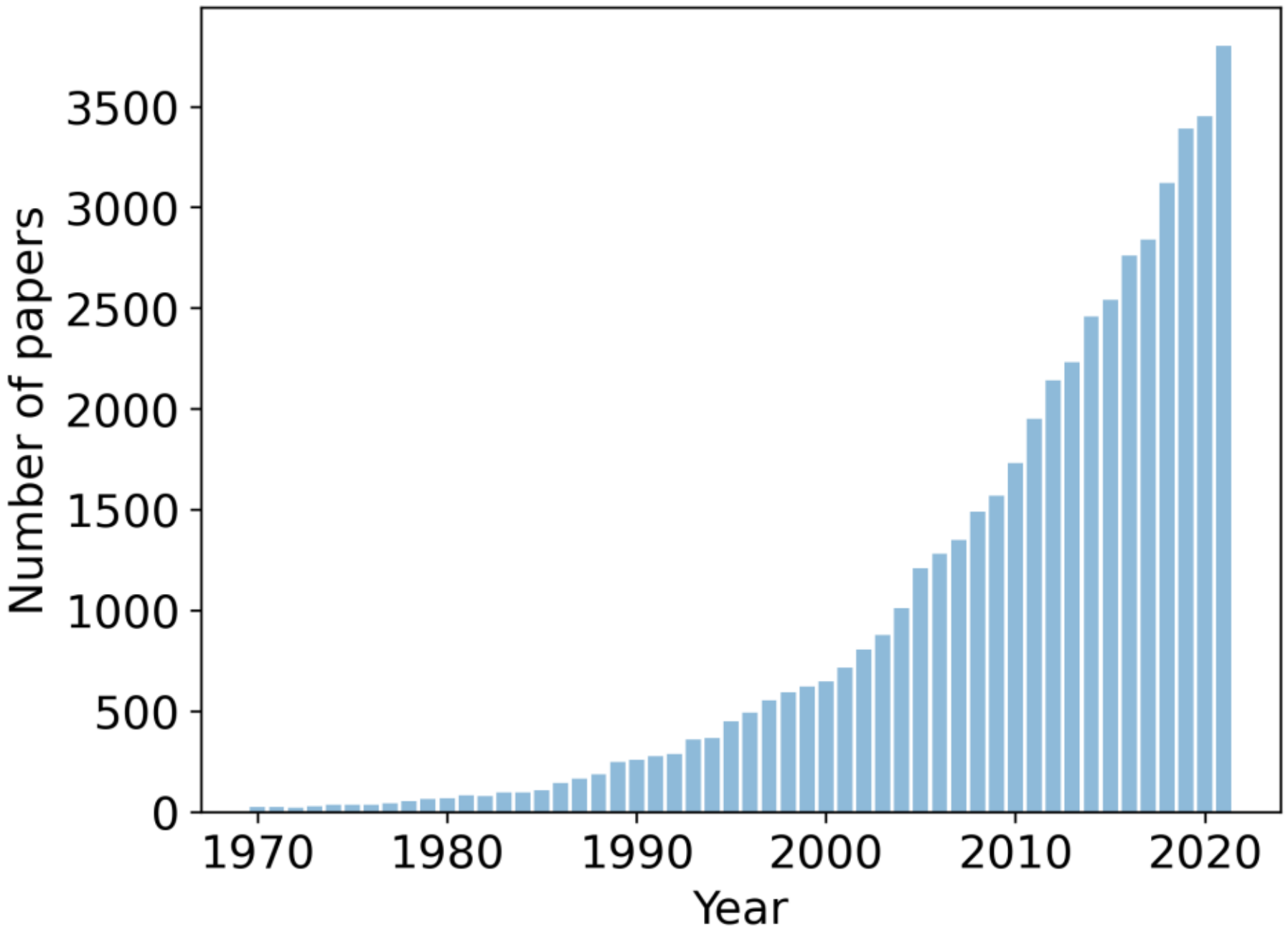

**Figure 1**. Number of papers which include the keywords “molecular dynamics” or “Monte Carlo” and either “stress tensor” or “pressure tensor” in the last 50 years, obtained from Google Scholar.[58]

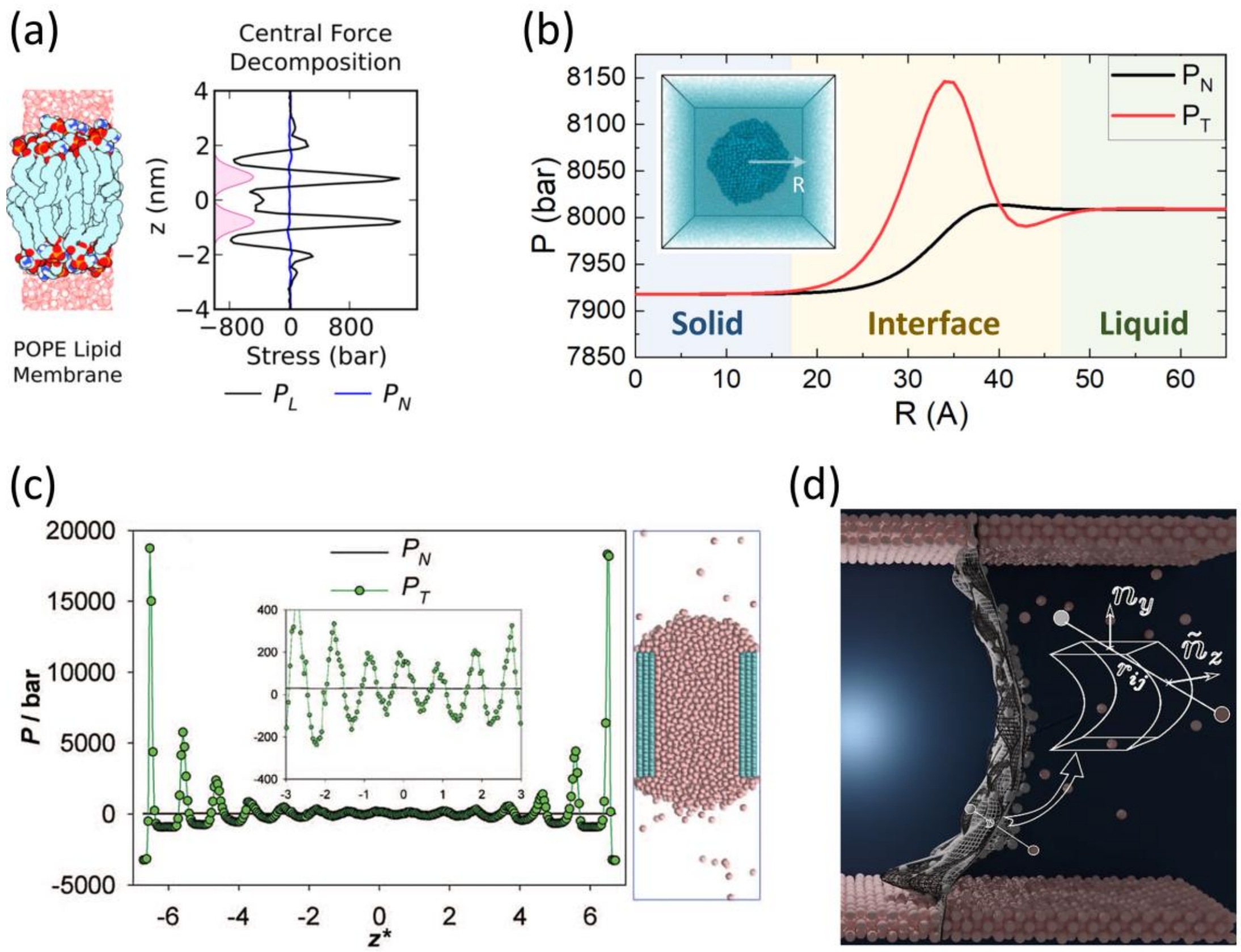

**Figure 2**. Illustration of applications for the microscopic pressure/stress tensor. (a) Local lateral ($P_L$) and normal stress ($P_N$) field in lipid membrane showing strong lipid unsaturation in the tail group regions. Adapted with permission from Ref. [7]. Copyright 2014 American Chemical Society. (b) Tangential ($P_T$) and normal ($P_N$) pressure profile in a solid-liquid nucleation system, showing strikingly lower pressure in the solid nucleus than the liquid environment. Data taken from Ref. [15]. (c) Snapshots and tangential and normal pressure profiles of Lennard-Jones argon (pink) confined in an atomistic carbon slit pore (cyan) at 87.3 K and 1 bar bulk pressure with pore width of 51 Å. High tangential pressure near the surface indicates the strong compression effect inside the physisorbed layers, shedding light on the understanding of high-pressure phenomena in more complex systems. Adapted from Ref. [34], with the permission of AIP Publishing. (d) The liquid-vapor-solid moving contact line: molecules from a molecular dynamics simulation shown with a Chebyshev function fitted to the liquid-vapor interface to be used in the stress calculation. The setup is a liquid bridge between two sliding molecular walls,[59] where the interface is split into bilinear patches and the stress is obtained as the force acting over each patch of the area.[60]

In the applications cited above, the range of length and time of interest can be very different, and we comment on the effect of these differences in later sections of this paper. We treat equilibrium and non-equilibrium systems separately in what follows. The paper is organized as follows. In Section 2, we introduce the fundamental pointwise forms of the pressure tensor. In Section 3, we describe the local pressure tensor for inhomogeneous systems that are at thermodynamic equilibrium. We introduce a thermodynamic route to the local pressure tensor, alternative to the common mechanical route. In Section 4, we consider the definition of the pressure tensor for non-equilibrium systems, where $\mathbf{P}$ depends on time as well as position in space. In Section 5, we point out several challenges associated with the microscopic pressure tensor, along with our perspectives on future developments. In Section 5.1, the historical controversies over the microscopic pressure tensor are discussed in detail. These include the non-uniqueness of the local pressure tensor due to the arbitrary contour integral, the possibility of defining a unique coarse-grained pressure through integration of the local pressure over some spatial domain, and the questions over the existence of the kinetic term in the stress tensor. In this review, we focus mainly on the formalisms of the local pressure/stress tensor for discrete particles that interact with short-range, isotropic, pairwise potentials. Extensions to complex molecular and material systems interacting with many-body and long-range potentials are discussed in Section 5.2. Other practical aspects that are also examined include current availability of software and computational tools (Section 5.3) and the possible experimental routes to the validation of the microscopic pressure tensor (Section 5.4). Lastly, concluding remarks are provided.

## 2. Fundamental equations for the pointwise pressure tensor

The local pressure tensor for a particle (point-mass) system with both spatial and temporal dependence can be written as[1,2]

$$\mathbf{P}(\mathbf{r},t) = \mathbf{P}^K(\mathbf{r},t) + \mathbf{P}^C(\mathbf{r},t) \tag{2}$$

where $\mathbf{P}^K(\mathbf{r},t)$ and $\mathbf{P}^C(\mathbf{r},t)$ are kinetic and configurational contributions, respectively, a tensor field defined at any given point in space $\mathbf{r}$ and time $t$. The kinetic contribution describes the flux of momentum:

$$\mathbf{P}^K(\mathbf{r},t) = \langle \sum_{i=1}^{N} \frac{\mathbf{p}_i\mathbf{p}_i}{m_i} \delta(\mathbf{r}-\mathbf{r}_i) \rangle \tag{3}$$

where the angular bracket $\langle \dots \rangle$ denotes the ensemble average; $N$ is the total number of particles in the system; $\mathbf{p}_i$ is the momentum of the particle $i$ and $\mathbf{p}_i\mathbf{p}_i$ is an outer product; $m_i$ is the mass of particle $i$; $\delta(\mathbf{r}) = \delta(x)\delta(y)\delta(z)$ is the delta function for a vector position in a Cartesian coordinate system. Here the momentum may include a streaming component, $\mathbf{p}_i/m_i = \dot{\mathbf{r}}_i - \boldsymbol{u}$, due to the streaming velocity $\boldsymbol{u}$ at point $\mathbf{r}$ being non-zero, where $\dot{\mathbf{r}}_i$ is the derivative with respect to time of the particle position. The mechanical definition of the kinetic pressure tensor in Eq. (3) corresponds to the ideal gas contribution in an equilibrium system (see Section 3.1) but must be defined in terms of streaming velocity in a non-equilibrium system (see Section 4.3).

For notational simplicity, we assume pairwise interactions here (additional terms due to many-body interactions are considered later in Section 5.2). The configurational pressure $\mathbf{P}^C$ can be obtained from the tensor product of pair forces between particles, $\mathbf{F}_{ij}$, and the line of interactions $\mathbf{r}_{ij}$ ($\mathbf{r}_{ij} = \mathbf{r}_j - \mathbf{r}_i$):

$$\mathbf{P}^C(\mathbf{r},t) = \frac{1}{2}\langle\sum_{i,j}^{N}\mathbf{F}_{ij}\mathbf{r}_{ij}O_{ij}\delta(\mathbf{r}-\mathbf{r}_i)\rangle \quad (4)$$

where the pre-factor $1/2$ accounts for double counting and the notation $\sum_{i,j}^{N}\mathbf{F}_{ij} = \sum_{i=1}^{N}\sum_{j\neq i}^{N}\mathbf{F}_{ij}$ has been introduced as shorthand for the double summation over all pairs of interactions. The $O_{ij}$ term is known as the Irving-Kirkwood (IK) operator,[1,61]

$$O_{ij} = 1 - \frac{1}{2!}\mathbf{r}_{ij}\cdot\frac{\partial}{\partial\mathbf{r}} + \cdots + \frac{1}{n!}(-\mathbf{r}_{ij}\cdot\frac{\partial}{\partial\mathbf{r}})^{n-1} + \cdots \quad (5)$$

which is obtained as the Taylor expansion in space of the difference between two Dirac delta functions for molecule $i$ and $j$,

$$\delta(\mathbf{r}-\mathbf{r}_i) - \delta\left(\mathbf{r}-\mathbf{r}_j\right) = -\mathbf{r}_{ij}\cdot\frac{\partial}{\partial\mathbf{r}}O_{ij}\delta(\mathbf{r}-\mathbf{r}_i) \quad (6)$$

If the expansion in Eq. (5) is simply truncated at the zeroth order term, $O_{ij} = 1$, we reach the so-called IK1 expression for the configurational part of the pressure tensor:

$$\overset{\text{IK1}}{\mathbf{P}}{}^{C}(\mathbf{r},t) = \frac{1}{2}\langle\sum_{i,j}^{N}\mathbf{F}_{ij}\,\mathbf{r}_{ij}\delta(\mathbf{r}-\mathbf{r}_i)\rangle \quad (7)$$

This is the *virial* pressure applied locally at a point in space.[62] For bulk homogeneous fluids where density is uniform throughout the system, the IK1 approximation ($O_{ij} = 1$) is exact.[61] For inhomogeneous fluids, such as those confined in nanopores or near interfaces, the IK1 approximation violates the mechanical equilibrium condition,[63,64] and will lead to erroneous results as interactions with the surrounding fluids are not included.[65] An exact form for the configurational pressure tensor can be reached by rewriting the IK operator using the fundamental theorem of contour integration:[2,66]

$$\mathbf{r}_{ij}O_{ij}\delta(\mathbf{r}-\mathbf{r}_i) = \oint_{C_{ij}}\delta(\mathbf{r}-\boldsymbol{\ell})d\boldsymbol{\ell} \quad (8)$$

where $C_{ij}$ denotes an arbitrary contour from the particle $i$ to the particle $j$, and $\boldsymbol{\ell}$ is the corresponding contour vector. Substituting Eq. (8) into Eq. (4) we arrive at the contour form of the configurational pressure tensor,[2]

$$\mathbf{P}^C(\mathbf{r},t) = \frac{1}{2}\langle\sum_{i,j}^{N}\mathbf{F}_{ij}\oint_{C_{ij}}\delta(\mathbf{r}-\boldsymbol{\ell})d\boldsymbol{\ell}\rangle \quad (9)$$

This equation is exact and allows the interaction between the molecules to be included at an arbitrary location unrelated to the molecules' positions.

While the kinetic pressure tensor (Eq. (3)) is well defined, the configurational part is non-unique due to the arbitrary contour involved in the calculation (Eq. (9)).[2] In practice, a particular contour definition needs to be chosen to arrive at an operational form of the local pressure tensor. By introducing a concept of surface element d$S$, Harasima[67] elegantly depicted two ways to assign a force contribution across such a surface of atomic dimension, which corresponds to two distinct definitions of the contour path. The first contour definition is a straight line of interactions between the molecules, known in the literature as the IK definition. It says that a pairwise force contributes to the pressure tensor at a surface element d$S$ if the joining straight line (contour) between two molecules passes through d$S$. This definition is consistent with Newton's assumption of impressed force between two points.[68,69] The second definition was first implicitly adopted by Kirkwood and Buff,[70] and later referred to as the Harasima definition. It says a pair-force contributes to the pressure tensor at the surface d$S$ if one of the molecules lies in the cylinder whose base is d$S$ (the axial direction of the cylinder is either parallel or perpendicular to the planar surface), and the other molecule is located on the other side of the plane of d$S$. Figure 3a and 3b illustrate the IK and Harasima definitions of the contour for a planar interface. Assuming the IK definition, the contour vector in Eq. (9) is simply $\boldsymbol{\ell} = \mathbf{r}_i + \lambda\mathbf{r}_{ij}$ with $0 \le \lambda \le 1$,[71] and Eq. (9) becomes

$$\overset{\text{IK}}{\mathbf{P}}{}^{C}(\mathbf{r},t) = \frac{1}{2}\langle \sum_{i,j}^{N} \mathbf{F}_{ij}\mathbf{r}_{ij} \int_0^1 \delta\big(\mathbf{r} - \mathbf{r}_i - \lambda\mathbf{r}_{ij}\big) d\lambda \rangle \tag{10}$$

Alternative but equally valid contour definitions are possible. Figures 3c and 3d show possible variations of the IK and Harasima contour definitions. In general, the IK contour definition is arguably the most convenient and natural choice. It can be readily implemented for a general three-dimensional (3D) pressure field[72] and in systems having an arbitrary geometry[8]. Compared to the other widely adopted contour choice of Harasima, the IK definition has been shown to be physically consistent in different coordinate systems (spherical[73] and cylindrical[74] coordinates). If long-range Coulombic interactions are present in the system, however, the Harasima contour is preferred due to its compatibility and optimal computational efficiency with the Ewald summation method.[6,74] No contour choice is more correct than the other in general. We will discuss this non-uniqueness problem further in detail in Section 5.1. For a bulk homogeneous system, the local pressure tensor is invariant to the choice of the contour definition. A spatial average of Eq. (9) over the entire system simplifies to the virial pressure tensor.[75]

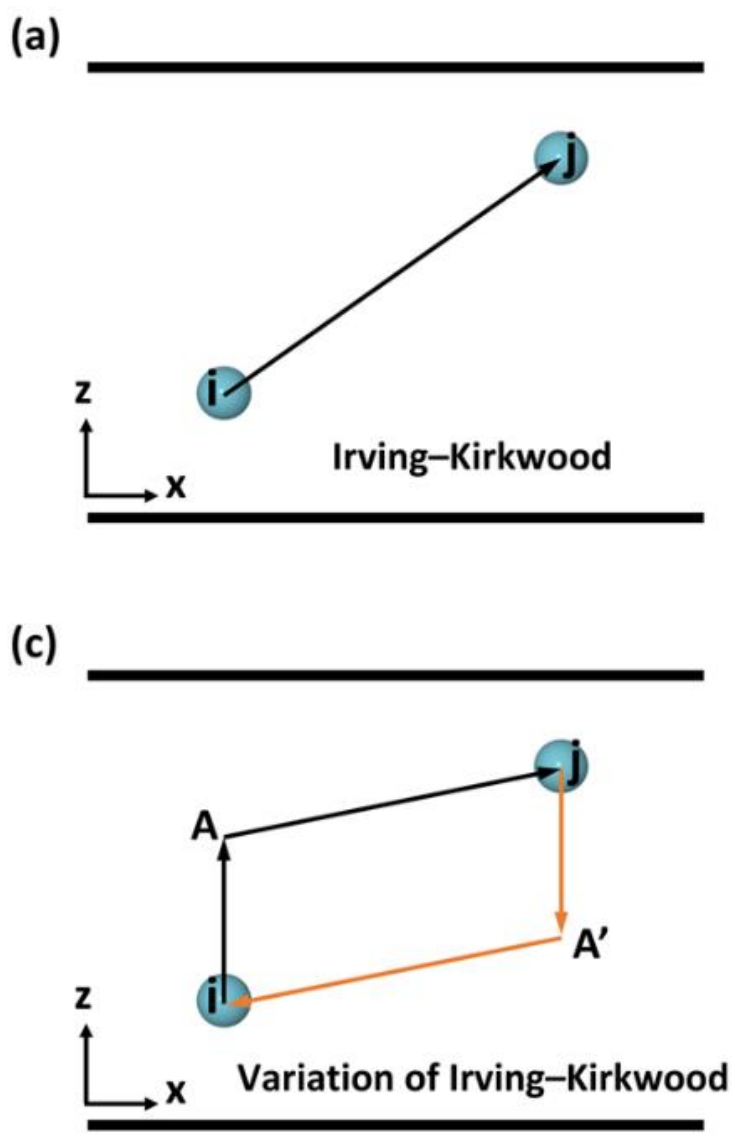

Figure 3. Four possible definitions of contour connecting the particles $i$ and $j$ for the local pressure tensor in a system of planar geometry. (a) Irving–Kirkwood (IK) definition. (b) Harasima (H) definition. (c) A variation of the Irving–Kirkwood (IK-VR) definition. (d) A variation of the Harasima (H-VR) definition. Due to the indistinguishability of particles, contours $C_{ij}$ (in black, from particles $i$ to $j$) and $C_{ji}$ (in orange, from particles $j$ to $i$) are equivalent and symmetric. Since $C_{ij}$ and $C_{ji}$ overlap for the IK and the H-VR definitions, only the contour $C_{ij}$ is plotted for clarity. All contours are projected onto the $xz$-plane, and the $z$-direction is perpendicular to the planar surface. These possible contour definitions illustrate the non-unique nature of the local pressure tensor. Reprinted from Ref. [35], with the permission of AIP Publishing.

## 3. Microscopic pressure tensor in equilibrium systems

For systems that are at thermodynamic equilibrium, the temporal dependence of the microscopic pressure tensor can be averaged out in the corresponding ensemble, and the spatial dependence of the pressure tensor is the main interest. Here we assume there is no shearing, as is the case for the equilibrium system, so that the off-diagonal elements in the pressure tensor simply vanish. In this section, we first discuss how the local pressure tensor is formulated in different geometries (coordinate systems) that are of practical importance for inhomogeneous systems at equilibrium. We then introduce a thermodynamic route to the microscopic pressure tensor. The equivalence between the thermodynamic route and the conventional mechanical (or virial) route is clarified.

### 3.1 Local pressure tensor in different geometries

**Planar geometry**. Systems that have planar interfaces are common and have been extensively investigated. Examples include fluids confined in a slit-shaped pore,[35] two phases (*e.g.*, gas and liquid) separated by a planar interface,[16] and planar self-assembled layers (*e.g.*, planar lipid bilayers).[6,7] The cartesian coordinate system is a convenient choice for planar geometries (Figure 4a). Here we take the $z$-axis to be the direction normal to the planar surface and assume homogeneity in the $xy$-plane, so that the local pressure tensor is a function of the $z$-position only. In the absence of external fields, the condition of mechanical (hydrostatic) equilibrium must be satisfied:[2]

$$\nabla \cdot \mathbf{P} = 0 \tag{11}$$

which has two implications:[76] 1) the tangential pressure $P_T$ has no local gradient in directions parallel to the $xy$-plane that induce flow anywhere; 2) the normal pressure $P_N$ should be a constant throughout the system:

$$P_N = P_{zz} = const. \tag{12}$$

That is for a two-phase system in equilibrium (phases $\alpha$ and $\beta$) separated by a planar interface, the normal pressure in the bulk phase $\alpha$ is equal throughout the interfacial region and into the bulk phase $\beta$ (Figure 5). For the slit-pore system, if we consider the material as a part of the system rather than an external field (no gravity is considered here), the normal pressure is constant across the entire pore.[35] It is worth noting that, due to the incommensurate packing of adsorbed layers in a slit pore, the normal pressure oscillates as the pore size increases.[33] The normal pressure converges to the pressure of the bulk phase that is in equilibrium with the adsorbed phase, in the limit of an open surface (*i.e.*, pore size is infinitely large).

The normal pressure is independent of the contour definition in Eq. (9).[35] By taking the normal ($zz$) component out of the pressure tensor in Eq. (9) and integrating (averaging) over the $x$ and $y$-directions (i.e., $xy$-surface), the normal pressure is given by,

$$P_N(z) = n(z)k_B T + \frac{1}{2S_z}\langle \sum_{i,j}^{N} \frac{{z_{ij}}^2}{r_{ij}} \frac{1}{|z_{ij}|} F_{ij} H\left(\frac{z - z_i}{z_{ij}}\right) H\left(\frac{z_j - z}{z_{ij}}\right)\rangle \tag{13}$$

where $n(z)$ is the local number density at $z$-position. The surface area is $S_z = L_x L_y$. If the sampling of the pressure tensor is carried out over the entire $xy$-plane, $L_x$ and $L_y$ will be the simulation box size in the $x$- and $y$-directions, respectively; $L_x$ and $L_y$ can also represent the size in the $x$- and $y$-directions for a specified region in the simulation box, if the pressure tensor is only sampled over that space.[35] The scalar $ij$-pair force is $F_{ij} = -\mathrm{d}u(r_{ij})/\mathrm{d}r_{ij}$, where $u(r_{ij})$ is the pair potential between particles $i$ and $j$ separated by a scalar distance $r_{ij}$. $z_{ij} = z_j - z_i$ is the $z$-component of vector $\mathbf{r}_{ij}$, and $H(x)$ is the Heaviside step function ($x > 0, H(x) = 1; x < 0, H(x) = 0; H(0) = 1/2$). We note that although the normal pressure is often written as a function of $z$ as in Eq. (13) for the use in molecular simulations, it is essentially a constant according to Eq. (12). The first term on the right-hand side of Eq. (13) is the kinetic (ideal gas) contribution to the pressure in equilibrium systems. It can be related to the mechanical definition in Eq. (3) at the limit of thermal equilibrium, by using the equipartition theorem,[77]

$$\frac{p_\alpha^2}{2m} = \frac{k_B T}{2} \tag{14}$$

where $p_\alpha$ is momentum of the molecule in $\alpha$-direction with $\alpha = x, y, z$[78] and we use $n(z) = \langle \sum_{i=1}^{N} \delta(z - z_i) \rangle$. We note that Eq. (14) is valid in any classical systems at thermal equilibrium, where the momenta and the coordinates are uncorrelated. For systems where quantum effects are significant, the equipartition theorem does not hold, and a quantum version of the theorem should be applied.[77] The second term in Eq. (13) is the configurational contribution arising from intermolecular interactions.

While the pressure normal to the $xy$-plane is well-defined, the tangential pressure parallel to the plane is not uniquely defined and depends on the intermolecular interaction contour definition. Because the system is homogeneous in the $xy$-plane, the local tangential pressure $P_T$ can be obtained by averaging over $P_{xx}$ and $P_{yy}$. Unlike the normal pressure, which is independent of $z$, the tangential pressure does depend on $z$. The local tangential pressure based on the IK contour definition (Eq. (10)) is given by[16,79]

$$\overset{\mathrm{IK}}{P_T}(z) = n(z) k_B T + \frac{1}{4S_z} \langle \sum_{i,j}^{N} \frac{{x_{ij}}^2 + {y_{ij}}^2}{r_{ij}} \frac{1}{|z_{ij}|} F_{ij} H\left(\frac{z - z_i}{z_{ij}}\right) H\left(\frac{z_j - z}{z_{ij}}\right) \rangle \tag{15}$$

Other equally valid contour definitions are possible. For example, assuming the Harasima (H) definition leads to:[16,67]

$$\overset{\mathrm{H}}{P_T}(z) = n(z) k_B T + \frac{1}{4S_z} \langle \sum_{i,j}^{N} \frac{{x_{ij}}^2 + {y_{ij}}^2}{r_{ij}} F_{ij} \delta(z - z_i) \rangle \tag{16}$$

where the Dirac delta function can be approximated as,

$$\delta(z - z_i) = \lim_{\Delta z \to 0} \frac{1}{\Delta z} \Lambda_z(z_i) = \lim_{\Delta z \to 0} \frac{1}{\Delta z} \left[ H\left(z_i - z + \frac{\Delta z}{2}\right) - H\left(z_i - z - \frac{\Delta z}{2}\right) \right] \tag{17}$$

where $\Lambda_z(z_i)$ is the so-called boxcar or top-hat function of $z_i$ for the interval from $z - \Delta z/2$ to $z + \Delta z/2$, which checks if $z_i$ is less than $z + \Delta z/2$ and greater than $z - \Delta z/2$. In practice, we choose $\Delta z \sim 0.001\sigma$ to numerically approximate the delta function where $\sigma$ is the Lennard-Jones (LJ) diameter of the particle. We note that using $\Lambda_z$ from Eq. (17) without taking the limit, Eq. (16) is equivalent to the 1D volume average (VA) form of the local pressure tensor. More discussions on the VA formalism will be presented in Section 4.4.

For gas-liquid interfaces, molecular simulation results show that the IK and Harasima definitions yield tangential pressures that differ by less than 10% (Figure 5).[16,80] For liquid-solid interfaces, the difference between these two contour definitions can be considerably larger.[34,35,63] The difference vanish when the local pressure tensor is integrated over the entire system (or at least over the inhomogeneous interfacial region), which is the case for the surface tension calculations.[16] The hydrostatic (averaged) pressure of the system can be calculated as the average of the trace of the pressure tensor, $P(z) = \left(P_{xx}(z) + P_{yy}(z) + P_{zz}(z)\right)/3$. Spatially averaging $P(z)$ over the entire system gives the bulk pressure, the configurational part of which is consistent with the virial theorem of Clausius.[81,82] This bulk pressure is unique and independent of the contour definition.

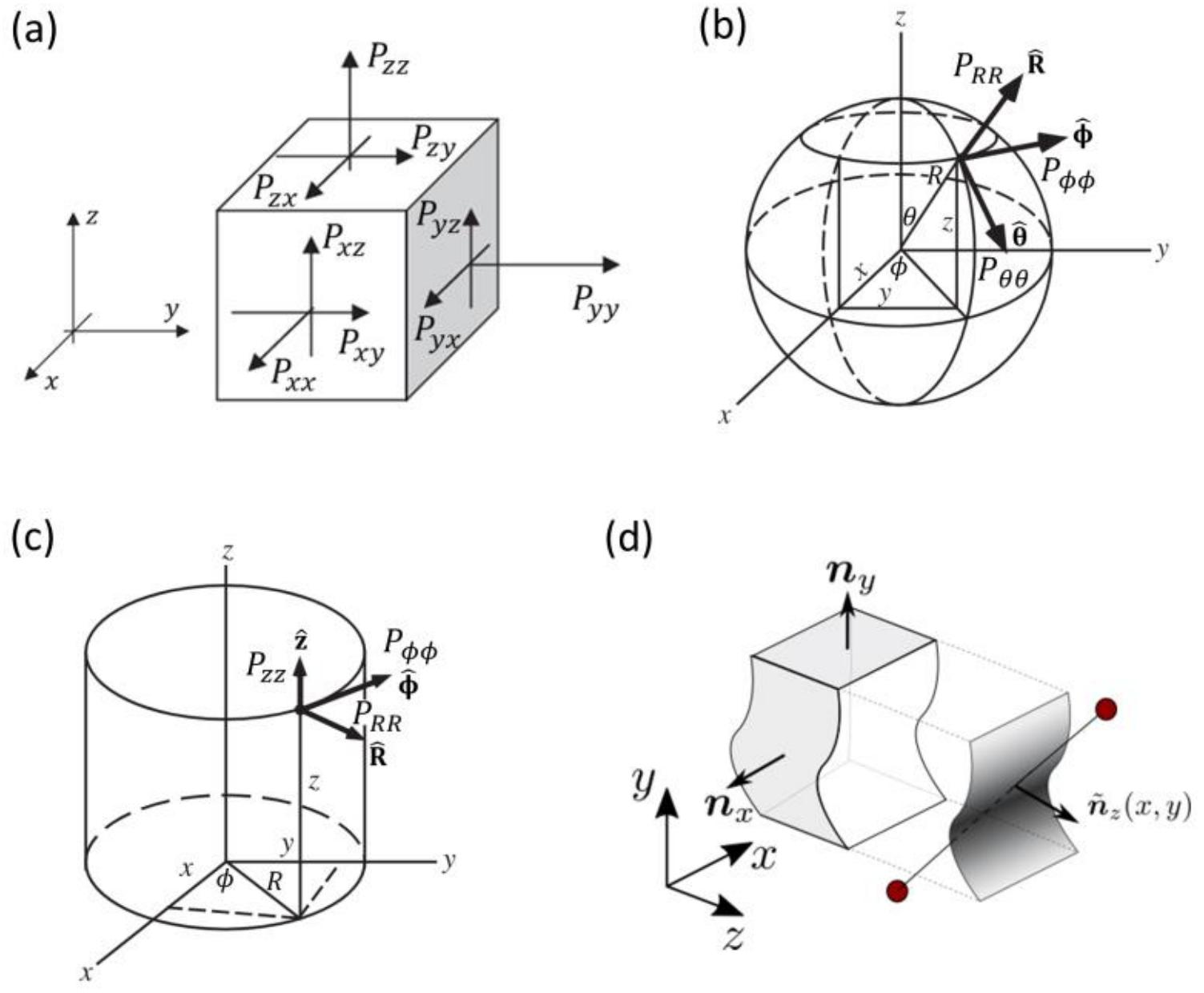

Figure 4. Pressure tensor definition in systems having different geometries. (a) Cartesian coordinates for a planar geometry. (b) Spherical coordinates for a spherical geometry. (c) Cylindrical coordinates for a cylindrical geometry. (d) An example of a more general coordinate system in terms of vector surface normal $\mathbf{n}_\alpha$ $(\alpha = x, y, z)$ where one (or more) surface is a general function $\tilde{\mathbf{n}}_z = \mathbf{n}_z(x, y)$, so that normal and tangential components of the pressure tensor will depend on surface position.

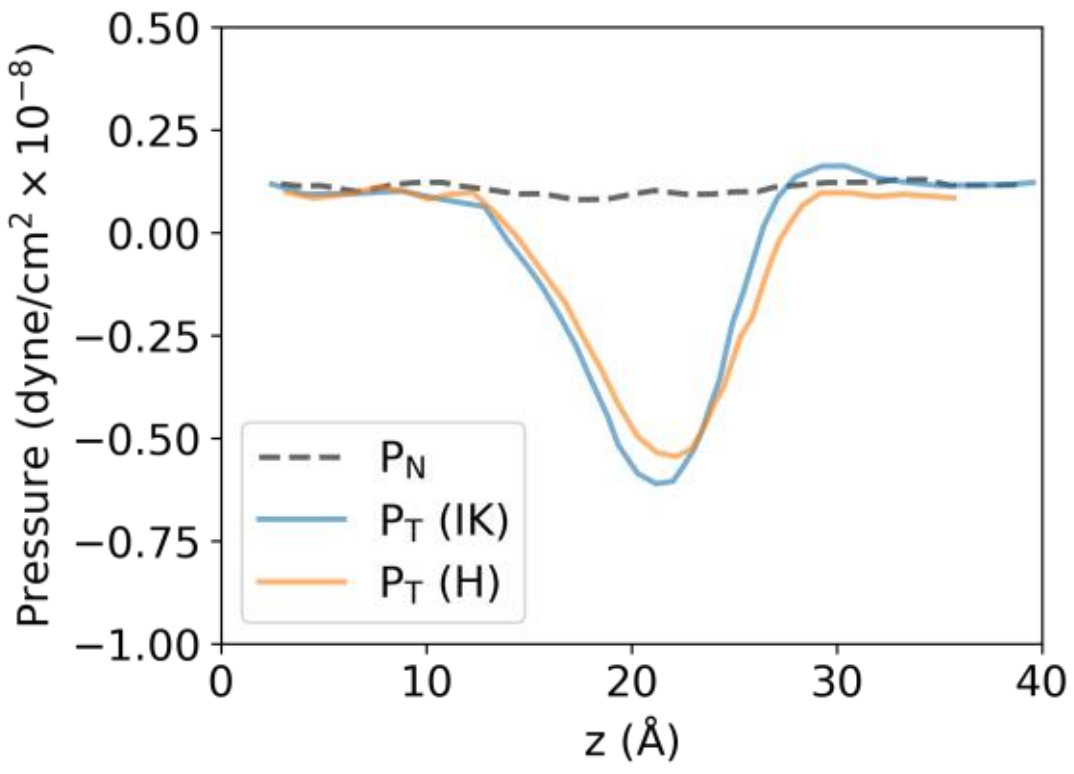

Figure 5. Pressure tensor profile across a planar gas-liquid interface for the argon-krypton mixture at $T = 115.77$ K, $N_{Ar}/N = 0.5$. These are smoothed molecular dynamics results for the truncated, shifted LJ

model, taken from Ref. [80]. The normal pressure $P_N$ is a constant (within statistical uncertainty) across the interface, and the tangential pressure $P_T$ by the IK and Harasima definitions are similar.

**Spherical geometry**. Examples of systems having a spherical geometry include gas/liquid/solid nuclei in the nucleation and crystallization process,[11–13,15,83] fluids confined in spherical pores,[84] and many spherical or quasi-spherical nanoparticles such as core-filled spherical nucleic acid.[85] Due to the symmetry of the system, it is convenient to calculate the local pressure tensor in spherical coordinates, $(R, \theta, \phi)$ (Figure 4b). The local spherical pressure tensor can be written as

$$\mathbf{P}(R) = P_{RR}(R)\hat{\mathbf{R}}\hat{\mathbf{R}} + P_{\theta\theta}(R)\hat{\boldsymbol{\theta}}\hat{\boldsymbol{\theta}} + P_{\phi\phi}(R)\hat{\boldsymbol{\phi}}\hat{\boldsymbol{\phi}} \tag{18}$$

where $P_{RR} = P_N$ is the normal pressure in the radial direction; $P_{\theta\theta}$ and $P_{\phi\phi}$ are the equivalent tangential components ($P_T$) due to symmetry in the polar and azimuthal directions, respectively; $\hat{\mathbf{R}}$, $\hat{\boldsymbol{\theta}}$, and $\hat{\boldsymbol{\phi}}$ are unit vectors that are orthogonal to each other. The mechanical equilibrium (Eq. (11)) dictates:

$$P_T(R) = P_N(R) + \frac{R}{2}\frac{dP_N(R)}{dR} \tag{19}$$

In spherical coordinates, the Harasima-like definition of the pressure tensor leads to unphysical results in the bulk system due to the singularity near the origin.[73,86] Therefore, the IK contour definition is commonly adopted in the literature. The normal (radial) pressure based on the IK contour definition is given by Ref. [87] (Detailed derivations for all components are available in the supporting information of Ref. [15]):

$$\overset{\text{IK}}{P}_{RR}(R) = n(R)k_BT + \frac{1}{8\pi R^2}\left\langle \sum_{i,j}^{N}\sum_{k=1}^{2}\frac{\left|\mathbf{r}_{ij}\cdot\hat{\mathbf{R}}_{\lambda_k}\right|}{r_{ij}}F_{ij}H(\lambda_k)H(1-\lambda_k)\right\rangle \tag{20}$$

where

$$\hat{\mathbf{R}}_{\lambda_k} = \begin{bmatrix}(x_i + \lambda_k x_{ij})/R\\(y_i + \lambda_k y_{ij})/R\\(z_i + \lambda_k z_{ij})/R\end{bmatrix} \tag{21}$$

and $\lambda_k$ are the roots of a quadratic equation, $(\mathbf{r}_{ij})^2\lambda^2 + 2\lambda\mathbf{r}_i\cdot\mathbf{r}_{ij} + (\mathbf{r}_i)^2 - R^2 = 0$. The polar pressure is given by[15]

$$\overset{\text{IK}}{P}_{\theta\theta}(R) = n(R)k_BT + \frac{1}{8\pi R^2}\langle\sum_{i,j}^{N}\sum_{k=1}^{2}\frac{\left(\mathbf{r}_{ij}\cdot\hat{\boldsymbol{\theta}}_{\lambda_k}\right)^2}{\left|\mathbf{r}_{ij}\cdot\hat{\mathbf{R}}_{\lambda_k}\right|}\frac{F_{ij}}{r_{ij}}H(\lambda_k)H(1-\lambda_k)\rangle \tag{22}$$

where

$$\hat{\boldsymbol{\theta}}_{\lambda_k} = \begin{bmatrix}\left(x_i + \lambda_k x_{ij}\right)\left(z_i + \lambda_k z_{ij}\right)/\left(Rd_{xy}\right)\\ \left(y_i + \lambda_k y_{ij}\right)\left(z_i + \lambda_k z_{ij}\right)/\left(Rd_{xy}\right)\\ -d_{xy}/R\end{bmatrix} \tag{23}$$

and $d_{xy} = \sqrt{\left(x_i + \lambda_k x_{ij}\right)^2 + \left(y_i + \lambda_k y_{ij}\right)^2}$. The azimuthal component is given by[15]

$$\overset{\text{IK}}{P}_{\phi\phi}(R) = n(R)k_BT + \frac{1}{8\pi R^2}\langle\sum_{i,j}^{N}\sum_{k=1}^{2}\frac{\left(\mathbf{r}_{ij}\cdot\hat{\boldsymbol{\phi}}_{\lambda_k}\right)^2}{\left|\mathbf{r}_{ij}\cdot\hat{\mathbf{R}}_{\lambda_k}\right|}\frac{F_{ij}}{r_{ij}}H(\lambda_k)H(1-\lambda_k)\rangle \tag{24}$$

where

$$\hat{\boldsymbol{\phi}}_{\lambda_k} = \begin{bmatrix}-\left(y_i + \lambda_k y_{ij}\right)/d_{xy}\\ \left(x_i + \lambda_k x_{ij}\right)/d_{xy}\\ 0\end{bmatrix} \tag{25}$$

In practice, to enhance the statistics, the tangential pressure is calculated as the average of the polar and azimuthal components, $P_T = \left(P_{\theta\theta} + P_{\phi\phi}\right)/2$.

**Cylindrical geometry**. Compared to the local pressure tensor for planar and spherical geometries, the theoretical development in a cylindrical geometry is generally overlooked. A complete derivation for the cylindrical pressure tensor was made available very recently.[21,74] The cylindrical pressure tensor is useful for understanding the behavior of systems having cylindrical interfaces. Examples include self-assembled micelles of a cylindrical shape,[88,89] a solid nucleus having a cylindrical shape,[90,91] and molecules confined in cylindrical pores.[74] Most of the synthetized porous materials with a well-defined geometry have cylindrical or quasi-cylindrical pores, such as carbon nanotubes and porous silica materials. The local cylindrical pressure tensor can be written as (Figure 4c)

$$\mathbf{P}(R) = P_{RR}(R)\hat{\mathbf{R}}\hat{\mathbf{R}} + P_{\phi\phi}(R)\hat{\boldsymbol{\phi}}\hat{\boldsymbol{\phi}} + P_{zz}(R)\hat{\mathbf{z}}\hat{\mathbf{z}} \tag{26}$$

where $P_{RR} = P_N$ is the normal pressure in the radial direction ($\hat{\mathbf{R}}$); $P_{\phi\phi}$ is the tangential pressure in the azimuthal direction ($\hat{\boldsymbol{\phi}}$); and $P_{zz}$ is the tangential pressure in the axial direction ($\hat{\mathbf{z}}$) with $P_{zz} \neq P_{\phi\phi}$. The mechanical equilibrium (Eq. (11)) dictates:

$$P_{\phi\phi}(R) = P_N(R) + R\frac{dP_N(R)}{dR} \quad (27)$$

Similar to the situation in spherical coordinates, the Harasima-like contour definition leads to a cylindrical pressure tensor with an unrealistic radial dependence near the origin in a bulk system.[74] In general, we suggest that any construction of the contour should not depend on polar coordinates (*i.e.*, radius and polar angles).[74] While a valid alternative to the Harasima-like definition in cylindrical coordinates is possible,[74] the IK contour definition naturally satisfies this polar-coordinate-independence condition, being a widely adopted choice of contour. The normal component of the cylindrical pressure tensor based on the IK contour definition is given by[21,74]

$$\overset{\text{IK}}{P}_{RR}(R) = n(R)k_BT + \frac{1}{4\pi R}\langle\sum_{i,j}^{N}\sum_{k=1}^{2}\frac{\left|\mathbf{r}_{ij}\cdot\widehat{\mathbf{R}}_{\lambda_k}\right|}{r_{ij}L}F_{ij}H(\lambda_k)H(1-\lambda_k)\rangle \quad (28)$$

where $L$ is the height of the cylinder. The unit radial vector is

$$\widehat{\mathbf{R}}_{\lambda_k} = \begin{bmatrix}\left(x_i+\lambda_k x_{ij}\right)/R\\ \left(y_i+\lambda_k y_{ij}\right)/R\\ 0\end{bmatrix} \quad (29)$$

and $\lambda_k$ are the roots of the equation $R^2 = \left(x_i+\lambda x_{ij}\right)^2 + \left(y_i+\lambda y_{ij}\right)^2$. The azimuthal component is written as[74]

$$\overset{\text{IK}}{P}_{\phi\phi}(R) = n(R)k_BT + \frac{1}{4\pi R}\langle\sum_{i,j}^{N}\sum_{k=1}^{2}\frac{\left(\mathbf{r}_{ij}\cdot\widehat{\boldsymbol{\Phi}}_{\lambda_k}\right)^2}{\left|\mathbf{r}_{ij}\cdot\widehat{\mathbf{R}}_{\lambda_k}\right|}\frac{F_{ij}}{r_{ij}L}H(\lambda_k)H(1-\lambda_k)\rangle \quad (30)$$

where

$$\widehat{\boldsymbol{\Phi}}_{\lambda_k} = \begin{pmatrix}-\left(y_i+\lambda_k y_{ij}\right)/R\\ \left(x_i+\lambda_k x_{ij}\right)/R\\ 0\end{pmatrix} \quad (31)$$

And the axial component of the cylindrical pressure tensor is given by

$$\overset{\text{IK}}{P}_{zz}(R) = n(R)k_BT + \frac{1}{4\pi R}\langle\sum_{i,j}^{N}\sum_{k=1}^{2}\frac{z_{ij}^2}{\left|\mathbf{r}_{ij}\cdot\widehat{\mathbf{R}}_{\lambda_k}\right|}\frac{F_{ij}}{r_{ij}L}H(\lambda_k)H(1-\lambda_k)\rangle \quad (32)$$

We note that Eqs. (30) and (32) are equivalent to the corresponding ones in Ref. [21] despite the difference in the appearance. Unlike equations in Ref. [21] that should be averaged over a number of $\phi$ and $z$ values in the simulation for the azimuthal and axial pressure, respectively, here we have already averaged over all possible $\phi$ and $z$ analytically through integration.[74]

**Arbitrary geometry**. While many systems of practical interest can be approximated with the three aforementioned simple geometries (planar, spherical, and cylindrical), there are still cases where systems have complex shapes without well-defined symmetries, and special treatments are needed. Typical methods for handling an arbitrary geometry involve the discretization of the system into local micro-volumes of molecular dimensions, and the local pressure tensor is evaluated either on the surface of the local volume[66] (Figure 4d) or as a spatial average over such local space[8,72]. A more general form of this kind is discussed in Section 4.4 for non-equilibrium systems.

## 3.2 Thermodynamic route to the pressure tensor and its equivalence to the mechanical route in thermodynamic equilibrium

So far, we have only talked about the mechanical route to the local pressure tensor, which is derived from the concept of "the force acting across a surface element d$S$". Equivalently, the pressure tensor can also be derived from a thermodynamic definition. It is instructive to start with the definition of the bulk (scalar) pressure in a canonical (NVT) ensemble (generalization to other ensembles is straightforward):

$$P = -\left(\frac{\partial A}{\partial V}\right)_{N,T} \tag{33}$$

where $A = -k_B T \ln Q$ is the Helmholtz free energy of the system with $N$ particles and a volume $V$ at temperature $T$. The canonical partition function $Q$ is defined as[82]

$$Q = \frac{1}{\Lambda^{3N} N!} \int \exp[-\beta \mathcal{U}(\mathbf{r}^N)] \mathrm{d}\mathbf{r}^N \tag{34}$$

where $\mathbf{r}^N \equiv \mathbf{r}_1, \mathbf{r}_2, \dots, \mathbf{r}_N$ represent the positions of all particles in the system; $\Lambda$ is the de Broglie wavelength; $\mathcal{U}$ is the total configurational energy of the system, and $\beta = 1/k_B T$. Following the pioneering work of Eppenga and Frenkel[92] on hard-core particles and Panagiotopoulos *et al.*[93] on systems with continuous potentials, the bulk pressure can be computed by considering a volume perturbation (VP) from $V$ to $V' = V + \Delta V$ (particle coordinates are scaled accordingly), with $\Delta V > 0$ being an infinitesimal, *isotropic* change of the volume, and Eq. (33) becomes[93]

$$\begin{aligned} \overset{\mathrm{VP}}{P} &\approx -\frac{A(V+\Delta V) - A(V)}{\Delta V} \\ &= \frac{k_B T}{\Delta V} \ln \left\langle \left(1 + \frac{\Delta V}{V}\right)^N \exp(-\beta \Delta \mathcal{U}) \right\rangle_V \end{aligned} \tag{35}$$

where $\Delta\mathcal{U} = \mathcal{U}(V+\Delta V) - \mathcal{U}(V)$ is the energy associated with the increase in volume, and the angular bracket with subscript $V$ denote a configurational average in the canonical ensemble over the unperturbed system of volume $V$. We note that Eq. (35) includes both kinetic and configurational contributions to the pressure. Eq. (35) also assumes the volume change is positive; in practice, a central finite-difference approximation with both positive and negative volume changes is recommended for better statistics.[94] Eq. (35) is the VP method or thermodynamic route to the pressure. A similar perturbation scheme for the surface (test-area method) has also been developed to compute the surface tension.[95–97] This test-area method is advantageous over the conventional pressure-tensor route to the surface tension, especially for small drops, because it can capture the entropic contributions due to the fluctuations in the energy of deformation.[98]

Similarly, Eq. (33) can be re-written to represent the diagonal Cartesian components of the pressure tensor $P_{\alpha\alpha}$ $(\alpha = x, y, z)$:[94]

$$P_{\alpha\alpha} = -\left(\frac{\partial A}{\partial V}\right)_{N,T,L_{\beta\neq\alpha}} \tag{36}$$

In this case, instead of performing an *isotropic* change of the volume as for the bulk pressure, only the simulation box dimension $L_\alpha$ in the $\alpha$-direction is perturbed to $L_\alpha + \Delta L_\alpha$ while all other dimensions $L_\beta$ $(\beta \neq \alpha)$ are kept fixed. Now we apply a common planar symmetry, *i.e.*, the $z$-axis is normal to the planar surface and the surface lies in the $xy$-plane, and we assume Eq. (36) is locally valid (*i.e.*, $Q$, $A$, and $V$ can be localized to a thin slab at $z$). By taking the partial derivative in Eq. (36) exactly, we can derive a local form of the pressure tensor:[34,94,99]

$$\begin{aligned} \overset{\mathrm{VP}}{P}_{\alpha\alpha}(z) &= n(z)k_BT - \left\langle\frac{\partial\mathcal{U}(z)}{\partial V(z)}\right\rangle_{N,T,L_{\beta\neq\alpha}} \\ &\approx n(z)k_BT + \frac{k_BT}{\Delta V(z)}\ln\left\langle\exp\big(-\beta\Delta\mathcal{U}(z)\big)\right\rangle_V \end{aligned} \tag{37}$$

where $\mathcal{U}(z)$ is the total configurational energy in an infinitesimally thin slab at a $z$-position, and $V(z) = L_xL_y\Delta z$ is the corresponding volume of the thin slab of width $\Delta z$. The first term on the right of Eq. (37) is the kinetic contribution, and the second term is the configurational contribution due to intermolecular interactions. The second line in Eq. (37) is a VP approximation similar to Eq. (35), but here the infinitesimal virtual expansion is performed only in the $\alpha$-direction, and $\Delta V(z) = S_\alpha\Delta L_\alpha\Delta z/L_z = S_\alpha\zeta L_\alpha\Delta z/L_z$, where $S_\alpha$ is the surface area normal to the $\alpha$-direction and $\zeta$ is a positive, infinitesimal number. In practice, the coordinates of all particles in the system (including those not in the thin slab at $z$) should be scaled by $\zeta$. $\Delta\mathcal{U}(z)$ is the change of the total potential energy in the thin slab at $z$ due to the corresponding virtual volume expansion. For confined fluids, for example, the total potential energy change should include contributions from both fluid-fluid and fluid-solid interactions.

Different criteria of assigning potential energy in slabs will lead to different local pressure tensor values,[34,100] again reflecting the non-uniqueness of the configurational contribution to the local pressure tensor. The ambiguity associated with the localization of the potential energy was also realized by Irving and Kirkwood.[1] Figure 6 illustrates two ways of assigning a pair interaction potential in space, labelled methods A and B. Method A assigns half of the pair potential energy, $u(r_{ij})/2$, into the thin slab that contains particle $i$, and the other half into the slab that contains particle $j$:[34,101]

$$\mathcal{U}(z) = \frac{1}{2}\sum_{i,j}^{N} u(r_{ij})\left[H\left(z_i - z + \frac{\Delta z}{2}\right) - H\left(z_i - z - \frac{\Delta z}{2}\right)\right] \tag{38}$$

Using Eqs. (17) and (38), it is straightforward to show that the first line of Eq. (37) (which is exact) for pairwise interactions will lead to the Harasima definition of the local tangential pressure in Eq. (16) by taking $\Delta z \to 0$ (see Ref. 74 for similar derivations in cylindrical coordinates). Method B equally distributes the pair potential $u(r_{ij})$ into slabs between two interacting particles $i$ and $j$:[34]

$$\mathcal{U}(z) = \frac{1}{2}\sum_{i,j}^{N} \frac{\Delta z}{|z_{ij}|} u(r_{ij}) H\left(\frac{z - z_i}{z_{ij}}\right) H\left(\frac{z_j - z}{z_{ij}}\right) \tag{39}$$

This energy attribution rule corresponds to the IK contour definition for the local pressure tensor (Eq. (15)). Similar to the integral contour whose choice is restricted in polar coordinates (for example, Harasima contour leads to invalid results in polar coordinates[73,74]), the ways to assign the local energy should also be regulated under certain conditions. Further studies are required to elucidate such restrictions. Finally, we note that the non-uniqueness in assigning the local energy may also explain the non-unique nature of the local heat flux.[102,103]

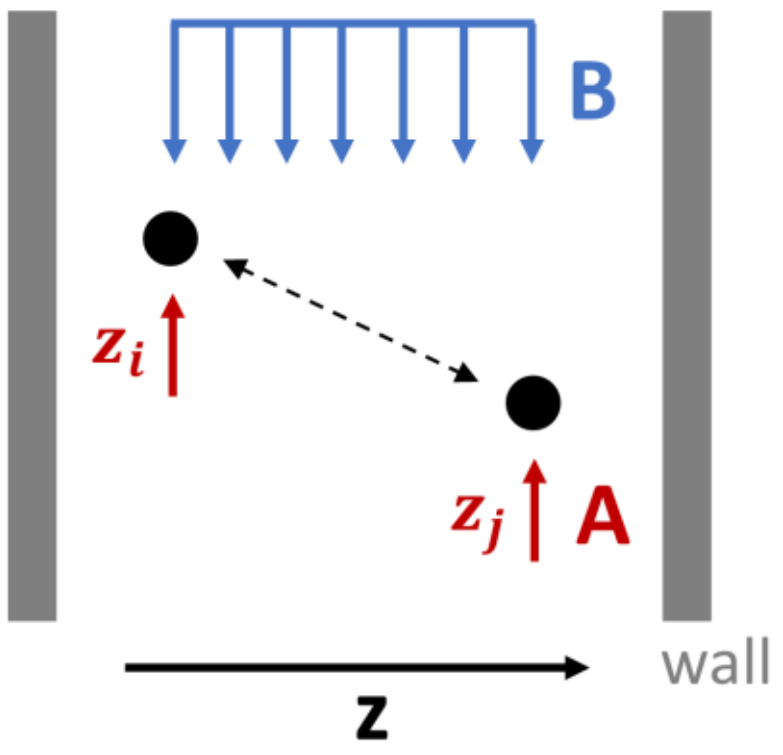

Figure 6. A schematic diagram showing the non-uniqueness involved in the distribution of potential energy in space for a slit-shaped pore system. A: half of the pair potential contributes to the thin slab containing particle $i$, and the other half contributes to the slab containing particle $j$. B: pair potential is equally distributed in the space between two interacting particles. This illustration reflects the non-uniqueness of the local pressure tensor from a thermodynamic perspective.

Extension of the thermodynamic route to curved interfaces, such as spherical[76] and cylindrical shapes[21,76], is possible. However, caution should be exercised due to the conjugate nature of polar components in the pressure tensor. For example, in spherical coordinates, perturbing the $R$-coordinates will *not* result in radial pressure because such rescaling also leads to perturbations in $\phi$- and $\theta$-directions. We will illustrate this point as follows. Let's assume we perform a small perturbation in the radial direction of a sphere having a radius of $R_0$, $R_0' = (1+\zeta)R_0$, where $\zeta$ is a positive, infinitesimal number. The volume change of the entire system is

$$\Delta V = V' - V = 3\zeta V \tag{40}$$

where in the last step we omit higher order terms involving $\zeta^2$ and $\zeta^3$. By perturbing the radius, the reversible work (free energy change) is done in all three polar directions:

$$\begin{aligned} \Delta A &= -\left(P_{RR} + P_{\theta\theta} + P_{\phi\phi}\right) \int_0^{R_0} 4\pi r^2 \zeta \mathrm{d}r \\ &= -\left(P_{RR} + P_{\theta\theta} + P_{\phi\phi}\right)\zeta V \end{aligned} \tag{41}$$

The integral in Eq. (41) is carried out over the entire sphere because the reversible work is done on the entire system (*i.e.*, all $R$-positions are rescaled with $\zeta$). Using Eqs. (40) and (41), and the thermodynamic definition of the pressure in Eq. (33), we get

$$\begin{aligned} P &= -\frac{1}{4\pi R^2}\left(\frac{\partial A}{\partial R}\right)_{N,T} \\ &= \frac{(P_N + 2P_T)}{3} \end{aligned} \tag{42}$$

where $P_{RR} = P_N$ is the normal pressure and $P_{\theta\theta} = P_{\phi\phi} = P_T$ is the tangential pressure due to symmetry. Eq. (42) clearly shows that perturbing the radial direction leads to a hydrostatic (bulk) pressure of the system defined as an average of the trace of the spherical pressure tensor. To decouple the normal and tangential pressure, we take advantage of the mechanical equilibrium condition in Eq. (19). Interested readers should refer to Ref. 76 for derivations of the local version of Eq. (42).

In short, the thermodynamic route (VP method) gives an equivalent form of pressure tensor to the one obtained from the mechanical route for equilibrium fluid systems (with off-diagonal elements being zero). Molecular simulation results indeed confirm this equivalence.[34,76,94,100] While both of them are useful for the calculation of the pressure tensor profile in inhomogeneous systems, they each have strengths and limitations. While the mechanical route is useful in studying the solid phase, the thermodynamic route is invalid in such a case. For a solid, the off-diagonal elements may not be zero, due to internal strain. That means if the solid is perturbated in one direction, as the thermodynamic route suggests, the results will be a coupling of the shear modes and the direct pressure. Nevertheless, the thermodynamic route is arguably more convenient for systems interacting with complex intermolecular potentials (*e.g.*, many-body interactions) where an explicit evaluation of the forces might be computationally challenging. Moving away from thermodynamic equilibrium, the mechanical route is the preferred choice, as discussed in the next section.

# 4. Microscopic pressure tensor in non-equilibrium systems

In this section, we extend the discussions of Section 3 to consider the definition of pressure with temporal evolution, convection, and flow in a moving reference frame. Moving away from equilibrium, we expect variation of the pressure tensor to drive flows, and for the pressure/stress tensor to depend on time as well as space (inhomogeneous). A simple example of this is a dynamic, or hydrodynamic, equilibrium in steady-state Couette flow, where the boundary conditions (or forcing/molecular walls) drive a flow giving a linear velocity profile and a constant shear stress. Many different computational techniques have been developed to study this system in non-equilibrium molecular dynamics (NEMD), including SLLOD,[104,105] Lees-Edwards boundary conditions,[106] and the application of shearing through tethered or fixed walls.[107,108] More generally, the onset of turbulence introduces convective terms which are non-zero even in an average sense. Finally, the most general case of non-equilibrium is an unsteady flow, where the gradient of convective transport and pressure together are equal to the time evolution of momentum in the system,

$$\underbrace{\frac{\partial \rho \boldsymbol{u}}{\partial t}}_{Time\ Evolution} + \underbrace{\boldsymbol{\nabla} \cdot \rho \boldsymbol{u}\boldsymbol{u}}_{Convection} = -\boldsymbol{\nabla} \cdot \mathbf{P} \tag{43}$$

where $\rho$ is the mass density and $\boldsymbol{u}$ is the streaming velocity. This time-evolving flow can occur in bubble growth, moving contact lines, onset of instabilities and many other areas of fluid dynamics.

We start by discussing the importance of temporal evolution on the definition of the stress tensor for a statistical mechanical ensemble in Section 4.1 before simplifying to the case of a single trajectory evolving in time in Section 4.2. We discuss the problem of defining kinetic pressure as we move away from equilibrium in Section 4.3. For the time evolution and convection, we need a clear mathematical framework for spatial localization of pressure valid away from equilibrium, and these are discussed in Section 4.4. In Section 4.5, we present recent works, and consider the most general case where the framework itself can move in time, which is useful in multi-phase fluid flow with deforming interfaces. We then outline a more pragmatic way to get this time evolution using mapping in Section 4.6. Finally, we discuss the statistical uncertainty involved in the pressure tensor calculations in Section 4.7.

## 4.1 Ensemble average and the time evolving phase space

Let $B(\mathbf{r}^N, \mathbf{p}^N)$ be some function of the $6N$ phase space variables. The ensemble average $\langle B \rangle$ is then the $6N$-dimensional integration over all position vectors and over all momenta vectors for an ensemble having a probability density function $f(\mathbf{r}^N, \mathbf{p}^N; t)$. Using the assumption that phase space is bounded,[1] and noting that for momentum $B = \sum_{i=1}^{N} m_i \dot{\mathbf{r}}_i \, \delta\big(\mathbf{r} - \mathbf{r}_i(t)\big)$, the time evolution of a phase-space averaged momentum is,

$$\frac{\partial}{\partial t}\langle\sum_{i=1}^{N} m_i \dot{\mathbf{r}}_i \, \delta\big(\mathbf{r} - \mathbf{r}_i(t)\big)\rangle = -\boldsymbol{\nabla}\cdot\langle \mathbf{P}^K + \mathbf{P}^C\rangle \tag{44}$$

where $\dot{\mathbf{r}}_i$ is the first-order derivative of the particle position with respect to time $t$. The right-hand side is the pressure introduced in Eqs. (3)-(4).

There are at least two purposes served by the ensemble average; the first is the practical reduction of noise, and the second is to ensure the validity of the Dirac delta function, which is not practically meaningful outside an integral (here the $6N$ dimensional integral of the ensemble average). The work of Noll[109] integrates the Dirac delta function over phase space and uses Noll's Lemma to address the $O_{ij}$ operator of Eq. (5), giving a form with similarities to the line integral of Eq. (8), but in Noll's phase-space integrated notation. The form is included in Appendix A2 and interested readers should refer to the review paper by Admal and Tadmor[78].

## 4.2 A single trajectory in time

If the ensemble average is dropped, equivalent forms of the pressure can still be obtained.[41] This step is essential in order to derive general equations free from the requirement of ensemble averaging, $\partial\langle B\rangle/\partial t \to \partial B/\partial t$. Here the form of $B = \sum_{i=1}^{N} m_i \dot{\mathbf{r}}_i \, \delta\big(\mathbf{r} - \mathbf{r}_i(t)\big)$ is as before. Applying the time derivative of momentum results, after some manipulation,[41,42] in the following,

$$\begin{aligned}
&\frac{\partial}{\partial t}\sum_{i=1}^{N} m_i \dot{\mathbf{r}}_i \, \delta\big(\mathbf{r} - \mathbf{r}_i(t)\big) \\
&= -\sum_{i=1}^{N} m_i \dot{\mathbf{r}}_i \dot{\mathbf{r}}_i \, \boldsymbol{\nabla}\cdot\delta\big(\mathbf{r} - \mathbf{r}_i(t)\big) + \sum_{i=1}^{N} m_i \ddot{\mathbf{r}}_i \delta\big(\mathbf{r} - \mathbf{r}_i(t)\big) \\
&= \ -\boldsymbol{\nabla}\cdot[\mathbf{P}^K + \mathbf{P}^C]
\end{aligned} \tag{45}$$

Here Newton's law, $\mathbf{F}_i = m_i \ddot{\mathbf{r}}_i$, is applied before expressing the force in a pairwise manner, and the delta functions are expanded as in the original work of Irving and Kirkwood.[1] The final right-hand side is the pressure at any instant, as in Eq. (44), but without the ensemble average. This has the advantage that it is a purely mechanical form of the pressure. More importantly, it is the starting point for a more general treatment.

However, we have introduced the formal mathematical problem to this pointwise pressure form that a Dirac delta function now exists outside an integral, which makes it poorly defined. To solve this problem, approaches in the literature approximate the delta function as a mollified weighting function[78,110–112] or approximate kernel[4,108,109]. A weighting function can be chosen to have any desired mathematical properties, such as compact support or normalization to unity. In this work, we do not approximate the delta function, instead we evaluate the formal integral of the delta function over a local volume in space. This control volume, or "finite volume" form[115] can be written in terms of surface tractions, and ensures the exact conservation of momentum during the single-trajectory time evolution, as shown in Section 4.4. Our treatment is equivalent to a choice of a uniform weighting function (a 3D boxcar function),[77] without any mollification. A weighting function may not satisfy momentum conservation, as it smears the momentum average over its functional form. This is analogous to the finite element method where different shape functions give different momentum distributions[117] and only the zeroth-order element (a finite volume) is conservative.[115] In the next section, the importance and difficulty of identifying and subtracting the hydrodynamic or streaming velocity is discussed, particularly when we move away from the ensemble picture.

## 4.3 Streaming velocity and the kinetic term

The dynamics of a fluid manifests itself through a velocity field $\boldsymbol{u}(\mathbf{r}, t)$ coupled with a scalar pressure $P(\mathbf{r}, t)$ at every point in space. This fluid velocity can be thought of as the average coherent motion of a stream of molecules, which we can define in terms of an instantaneous form of the Irving-Kirkwood momentum and density,

$$\boldsymbol{u}(\mathbf{r}, t) = \frac{\rho(\mathbf{r}, t)\, \boldsymbol{u}(\mathbf{r}, t)}{\rho(\mathbf{r}, t)} = \frac{\sum_{i=1}^{N} m_i \dot{\mathbf{r}}_i\, \delta(\mathbf{r} - \mathbf{r}_i)}{\sum_{i=1}^{N} m_i\, \delta(\mathbf{r} - \mathbf{r}_i)} \tag{46}$$

Here $\rho(\mathbf{r}, t)$ is the mass density of the fluid; $\boldsymbol{u}(\mathbf{r}, t)$ is the average velocity of the molecules, often known as the streaming velocity. The kinetic pressure can then be defined by introducing the peculiar velocity $\mathbf{p}_i/m_i$, the particle motion not contributing to the net velocity field,

$$\frac{\mathbf{p}_i}{m_i} = \dot{\mathbf{r}}_i - \boldsymbol{u}(\mathbf{r}, t) \tag{47}$$

A clear problem with the instant trajectory is apparent here: for a single timestep in a molecular simulation, there is no way to split fluctuation and streaming parts. Of a molecular kinetic motion, the contribution which becomes velocity and the contribution which is kinetic pressure can only be determined with both spatial or temporal averaging. We discuss the spatial averaging in Section 4.4. The length of temporal averaging will adjust measurements of velocity, and requires some pragmatism in choice. However, this does not mean that the use of a single trajectory is wrong, as will be discussed below. In fact, chaotic trajectory divergence means an ensemble approach is not always possible in situations away from an attractor state,[78,118] so piecewise temporal averaging may be the only approach to get velocity in highly non-linear systems.

Taking the kinetic term in Eq. (45), substituting in the peculiar velocity of Eq. (47) and applying the definition of momentum and density allow the total kinetic tensor to be written in terms of a kinetic pressure and a convection term,

$$\sum_{i=1}^{N} m_i \dot{\mathbf{r}}_i \dot{\mathbf{r}}_i \, \delta(\mathbf{r}-\mathbf{r}_i) = \sum_{i=1}^{N} \frac{\mathbf{p}_i \mathbf{p}_i}{m_i} \delta(\mathbf{r}-\mathbf{r}_i) + \rho(\mathbf{r},t)\boldsymbol{u}(\mathbf{r},t)\boldsymbol{u}(\mathbf{r},t) \tag{48}$$

In an equilibrium system, the second term on the right (convection) is equal to zero. In the case of hydrodynamic equilibrium, for example in channel flow, convection is also negligible, and the velocity varies in a well-defined way. This is true even in molecular systems down to 5-10 atomic diameters,[42] *e.g.* a linear function $u(y)\sim y$ in Couette flow[119] or parabolic in pressure-driven Poiseuille flow $u(y)\sim y^2$.[120] When the expected velocity profile is known, such as in Couette or Poiseuille flow, it makes the definition of peculiar velocity, $\mathbf{p}_i/m_i$, straightforward.[121] This idea of constructing quantities with a known velocity inspired some of the earliest non-equilibrium molecular dynamics (NEMD) simulation methods, by applying a known forcing function, so that the resulting velocity is as expected.[122] However, as molecular simulation increasingly pushes to more complex non-equilibrium cases, a well-defined velocity profile is no longer possible. Examples include complex flow patterns around obstacles,[123–125] rolling or cells formed by thermal gradients,[126,127] vortex formation,[128] Taylor Couette flow,[129,130] the Rayleigh-Taylor instability[131,132] or shock wave instability[133,134]. There is also a recent explosion in papers using molecular simulations in multi-phase flows for films, droplets, bubbles and contact lines. We defer a consideration of these to Section 4.5, where a moving interface reference frame is presented specifically for these types of problems. In all non-linear single-phase examples, the evolution of velocity is chaotic, which makes the concept of an ensemble averaging problematic, as different simulations would diverge given enough time.

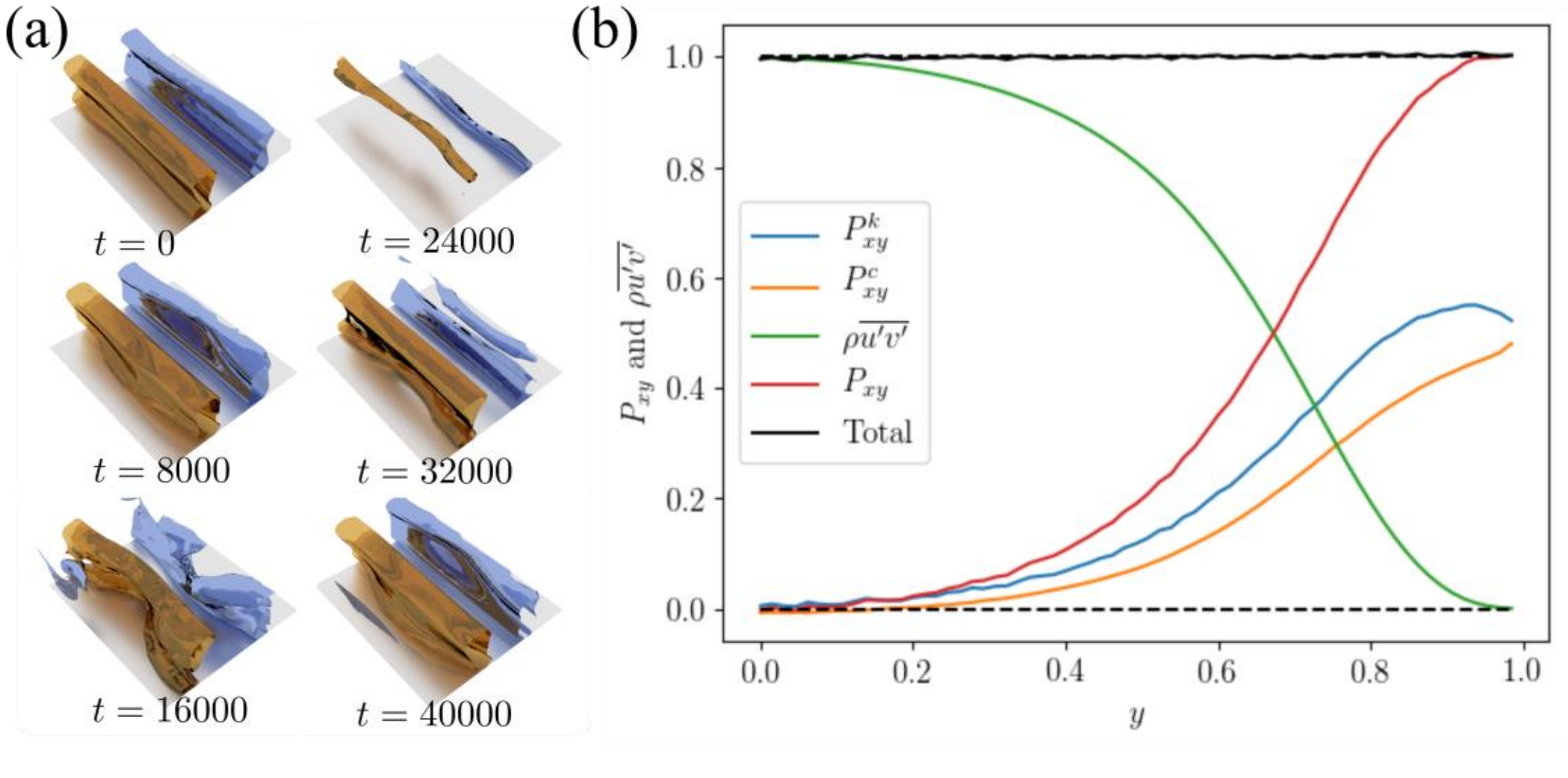

Figure 7. Turbulent Couette flow in a molecular simulation. (a) The evolution of iso-surfaces of turbulent kinetic energy (TKE) within a regeneration cycle with positive TKE in blue and negative in orange; (b) The average pressure moving across the channel from channel center to the top wall ($y = [-1,1]$ in the wall-normal direction), showing the contribution of kinetic pressure, $P^K_{xy}$, and configurational pressure, $P^C_{xy}$, which add to give total pressure, $P_{xy}$, and with the turbulent shear stress $\rho\overline{u'v'}$ give a constant (black line)

at all points in the channel (satisfying mechanical equilibrium). Note that all quantities are normalized by $\tau_0$, the shear stress between the wall and first layer of the fluid.

Perhaps the most interesting and general example of splitting the streaming velocity from the kinetic contribution is for turbulent flow.[135] This is shown in Figure 7 for the smallest known cases of time-steady turbulent flow[136,137] but simulated in a molecular dynamics simulation[138]. This minimal-channel Couette flow exists at a Reynolds number of 400, which requires $N \approx 300$ million molecules taking an average density of 0.3 in LJ units and temperature chosen to minimize viscosity. Convection is non-zero and velocity varies in time, so we use Reynolds decomposition, $\boldsymbol{u} = \bar{\boldsymbol{u}} + \boldsymbol{u}'$, to define a long-time-average streaming velocity $\bar{\boldsymbol{u}}$**.** This average velocity allows us to define a turbulent fluctuating part, denoted here by the prime $\boldsymbol{u}'$. This $\boldsymbol{u}'$ is like the peculiar velocity in NEMD, a velocity which is not contributing to the mean flow, but the fluctuations are large scale eddying motions instead of molecular fluctuations of $\mathbf{p}_i$. The typical cycle of fluctuations is shown in Figure 7a as the iso-surfaces of squared velocity $|\boldsymbol{u}'|^2 = u'^2 + v'^2 + w'^2$ where $u'$, $v'$ and $w'$ are the three velocity components. As with the kinetic pressure tensor, knowing the average velocity allows the average fluctuations to be identified, which is called the Reynolds stress tensor:

$$\rho\overline{\boldsymbol{u}'\boldsymbol{u}'} = \rho\boldsymbol{u}\boldsymbol{u} - \rho\bar{\boldsymbol{u}}\bar{\boldsymbol{u}} \tag{49}$$

This is a dimensionally and physically the same form as the kinetic pressure, the outer product of fluctuating velocity components, but on a larger scale. Instead of small molecular fluctuations, it is the average momentum carried by turbulent eddies which make up the Reynolds stress. Indeed, Osborne Reynolds apparently defined this stress, inspired by the subtraction of streaming velocity seen in his earlier work on kinetic theory.[139,140] The mean flow is time stationary on a longer time scale, with the flow going through a regeneration cycle, where the streaks breakdown the energizing flow vortices before a regeneration occurs (Figure 7a).This cycle repeats indefinitely and a long-time average can be collected. The average pressure and Reynolds shear stress in the top half of the symmetrical channel are shown in Figure 7b where the total shear contribution is constant, *i.e.*, $[\rho\overline{u'v'} + P_{xy}^K + P_{xy}^C]/\tau_0 = 1$ . Interestingly, the kinetic and configurational contributions to shear stress are similar in magnitude in Figure 7b. It is similar to the case of hydrodynamic stability in laminar Couette flow, where adding the average Reynolds shear stresses together with the shear pressure gives a constant, *i.e.* it satisfies $\boldsymbol{\nabla} \cdot [\mathbf{P} + \rho\overline{\boldsymbol{u}'\boldsymbol{u}'}] = 0$, and as a result $d\bar{\boldsymbol{u}}/dt = 0$. This turbulent equilibrium is a property of the relatively simple channel flow and would not be true in general.

The off-diagonal components of the Reynolds stress tensor in Eq. (49), e.g., $\rho\overline{u'v'}$ are like the shear stress $P_{xy}^K$ in that they include $x$-momentum carried in the $y$-direction. As a result, the kinetic pressure can be split into velocity existing on three length and time scales,

$$\langle\sum_{i=1}^{N} m_i \dot{\mathbf{r}}_i \dot{\mathbf{r}}_i \delta(\mathbf{r}-\mathbf{r}_i)\rangle = \underbrace{\langle\sum_{i=1}^{N} \frac{\mathbf{p}_i \mathbf{p}_i}{m_i} \delta(\mathbf{r}-\mathbf{r}_i)\rangle_t}_{\mathbf{P}^K} + \rho\overline{\boldsymbol{u}'\boldsymbol{u}'} + \rho\overline{\boldsymbol{u}\boldsymbol{u}} \tag{50}$$

Here we have included a time averaging, $\langle\ldots\rangle_t$, over a shorter timescale $t_{MD}$. The overbar denotes an average over the length of the simulation, in this case $t_{Total}$. Eq. (50) highlights a very interesting insight into the time averaging process; as the averaging period of $\langle\ldots\rangle_t$ increases relative to the overbar, the kinetic contribution will increasingly be assigned to molecular kinetic shear stress $\mathbf{P}^K$ and Reynolds stress will decrease, with the limit $t_{MD} \to t_{Total}$ seeing all turbulent fluctuations counted as kinetic pressure. In Eq. (50), we see kinetic pressure as simply the first order term in a series expansion of fluctuations at increasing length and time scales. This is a source of non-uniqueness in the kinetic pressure away from equilibrium, intimately linked to the turbulent cascade, where fluctuations $\boldsymbol{u}'$ form a continuous spectrum of scales. In the turbulence literature, understanding and modeling the range of turbulent scales is one of the main focuses of research.[141,142] This section shows that the uncertainty in splitting molecular motion into kinetic pressure and streaming velocity is part of a larger picture in fluid dynamics. Here flow is multi-scale over many orders of magnitude and care is needed to decide which scales must be modelled. In experimental Aerodynamics, the pressure measured by a pitot tube, which physically slows the flow velocity to zero, or stagnation, is known as the total pressure. This is to distinguish it from static pressure due only to kinetic collisions measured in a static flow. The future of NEMD simulation will need to link closely to the fluid community to address these problems.[141,142]

## 4.4 Localization of the pressure tensor in space

The field of NEMD closely mirrors fluid dynamics, where fluid properties are defined on a field, which is expressed as a discrete grid of values in CFD. For this reason, expressing the pressure on a tessellating grid of cells is a natural choice to explore fluid phenomena. This process starts by integrating the Irving-Kirkwood form of pressure over a volume $V$ in space, including the kinetic pressure of Eq. (3) and the configurational pressure of Eq. (10) using the IK contour,

$$\begin{aligned} &\int_V [\mathbf{P}^K(\mathbf{r},t) + \mathbf{P}^C(\mathbf{r},t)]\, dV \\ &= \sum_{i=1}^{N} \frac{\mathbf{p}_i \mathbf{p}_i}{m_i} \int_V \delta(\mathbf{r}-\mathbf{r}_i)\, dV + \frac{1}{2}\sum_{i,j}^{N} \mathbf{F}_{ij}\, \mathbf{r}_{ij} \int_V \int_0^1 \delta(\mathbf{r}-\mathbf{r}_i-\lambda\mathbf{r}_{ij}) d\lambda\, dV \\ &= \sum_{i=1}^{N} \frac{\mathbf{p}_i \mathbf{p}_i}{m_i} \vartheta_i + \frac{1}{2}\sum_{i,j}^{N} \mathbf{F}_{ij}\, \mathbf{r}_{ij} \int_0^1 \vartheta_\lambda d\lambda \end{aligned} \tag{51}$$

where $\vartheta_i$ and $\vartheta_\lambda$ are the integral of the Dirac delta functions over a finite volume, in the case of a cuboid centered at $\mathbf{r} = (x, y, z)$ and having a dimensional $\Delta\mathbf{r}$. The function $\vartheta_i$ can be written as $\vartheta_i = \Lambda_x(x_i)\Lambda_y(y_i)\Lambda_z(z_i)$, where $\Lambda_\alpha(\alpha_i)$ with $\alpha = x, y, z$ is the boxcar function introduced in Eq. (17), and $\vartheta_i = 1$ when the particle $i$ is inside a cuboid and $\vartheta_i = 0$ otherwise. Assuming a single average value of pressure in a volume, $\int_V \mathbf{P}dV = \overset{\text{VA}}{\mathbf{P}}\Delta V$, Eq. (51) results in the so-called Volume Average (VA) pressure tensor,

$$\overset{\text{VA}}{\mathbf{P}} = \frac{1}{\Delta V}\left[\sum_{i=1}^{N} \frac{\mathbf{p}_i\mathbf{p}_i}{m_i}\vartheta_i + \frac{1}{2}\sum_{i,j}^{N} \mathbf{F}_{ij}\,\mathbf{r}_{ij} l_{ij}\right] \tag{52}$$

where $\Delta V$ is the local volume, and the shorthand $l_{ij} = \int_0^1 \vartheta_\lambda d\lambda$ gives the fraction of the interaction contour $\boldsymbol{\ell}$ inside the averaging volume. We note that the streaming velocity $\boldsymbol{u}(\mathbf{r}, t)$, defined in Eq. (46), is averaged over the same volume in the VA form (Figure 8a), *i.e.*, $\boldsymbol{u}(\mathbf{r}, t) = \sum_{i=1}^{N} m_i \dot{\mathbf{r}}_i \vartheta_i / \sum_{i=1}^{N} m_i \vartheta_i$. This VA pressure was originally proposed in a 1D form for shockwaves by Hardy in 1982.[110] It was then extended by Cormier *et al.*[143] to a spherical volume and made more formal by Murdoch.[111,112] For the case of the IK contour, the fraction of line $l_{ij}$ can be obtained exactly in a cube from plane-line intersections shown in Figure 8b, or in a sphere[87,144] or cylinder[21] from surface-line intersections. For more complicated local volume shapes, $l_{ij}$ can be obtained by splitting the line into segments and binning each segment numerically if it is inside the volume. As an example, the Voronoi decomposition of Hatch and Debenedetti[8] is shown in Figure 8c. They developed a formalism that enables the calculation of the local stress tensor on an atom or an arbitrary group of atoms by averaging over the local volume of this group. The local volume for a certain group was obtained using the Voronoi decomposition method. In their formalism, the contour segment that is within the volume $V_g$ of the targeting group of atoms contributes to the local stress of this group. For instance, in Figure 8c, the red segment of the line connecting particles 3 and 4 contributes to the local stress of the group composed of particles 2 and 5. In this way, it is identical to the VA pressure of Eq. (52) which uses length of interaction inside the volume, but with the volumes chosen based on molecular structure. The presence of a molecule at the center of a volume changes the measured pressure to include insight into the material structure, with some similarity to the molecular centric radial distribution function.[144]

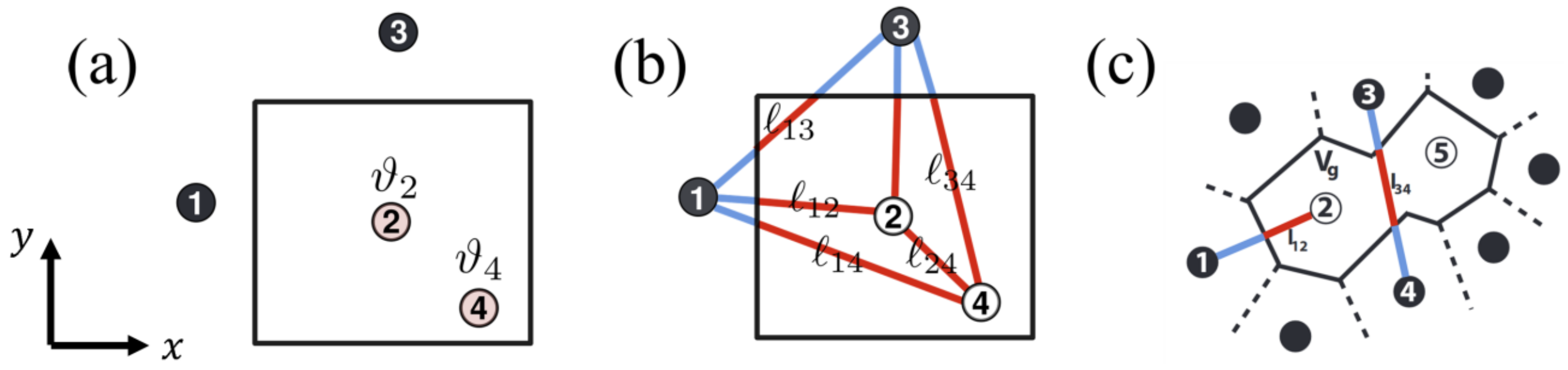

Figure 8. The volume average (VA) form of pressure tensor, showing (a) the molecules 2 and 4 are inside the local volume and contribute to the kinetic pressure; (b) the length of line $l_{ij}$ in the local volume used for the configurational pressure (contributions are shown in red); and (c) a more general implementation using Voronoi volumes, adapted from Ref. [8], with the permission of AIP Publishing. In both (b) and (c), the IK contour is used.

On the other hand, the pressure tensor can be expressed in a plane form. For example, by taking the three components of the VA pressure tensor in Eq. (52) which are acting on the surface that is normal to the $x$-direction, namely $\mathbf{P}_x = [P_{xx}, P_{xy}, P_{xz}]$, and evaluating the limit that the volume tends to zero in the $x$-direction, we arrive at[145]

$$\lim_{\Delta x\to 0} \overset{\text{VA}}{\mathbf{P}_x}(\mathbf{r},t) = \lim_{\Delta x\to 0}\frac{1}{\Delta x\Delta y\Delta z}\left[\sum_{i=1}^{N}\frac{\mathbf{p}_i p_{ix}}{m_i}\vartheta_i + \frac{1}{2}\sum_{i,j}^{N}\mathbf{F}_{ij}\, x_{ij} l_{ij}\right]$$
$$= \frac{1}{\Delta S_x}\left[\sum_{i=1}^{N}\frac{\mathbf{p}_i p_{ix}}{m_i}\delta(x-x_i)\,\Lambda_y(y_i)\Lambda_z(z_i)\right.$$
$$\left. + \frac{1}{2}\sum_{i,j}^{N}\mathbf{F}_{ij}\, x_{ij}\int_0^1 \delta\left(x - x_i - \lambda x_{ij}\right)\Lambda_y(y_\lambda)\Lambda_z(z_\lambda)d\lambda\right] \tag{53}$$

where $\Delta S_x = \Delta y \Delta z$ is the surface element normal to the $x$-direction. In the limiting case, Eq. (17) is used to convert a boxcar function to a delta function. If we take limits of $\Delta y$ and $\Delta z$ to be the edges of the simulation box with periodic boundaries, we have $\lim_{\Delta y\to L_y}\Lambda_y = \lim_{\Delta z\to L_z}\Lambda_z = 1$, $\Delta S_x = S_x$, and Eq. (53) simplifies to the method of planes (MoP) pressure:[61]

$$\overset{\text{MoP}}{\mathbf{P}}_x(x) = \frac{1}{S_x}\left[\sum_{i=1}^{N}\frac{\mathbf{p}_i p_{ix}}{m_i}\delta\left(x - x_i(t)\right)\right.$$
$$\left. + \frac{1}{4}\sum_{i,j}^{N}\mathbf{F}_{ij}\left[sgn(x-x_i) - sgn\left(x - x_j\right)\right]\right] \tag{54}$$

The integral along $\lambda$ has been evaluated to give signum functions (for a contour, use $\oint_{x_i}^{x_j}\delta(x-\ell_x)f(\ell_x)d\ell_x = f(x)\left[H(x-x_i) - H\left(x-x_j\right)\right]$), where $sgn(x) = 1$ for $x > 0$ and $sgn(x) = -1$ for $x < 0$, and $sgn(x) = 0$ otherwise. The interpretation of the differences in signum functions is that $x_i$ and $x_j$ have to be on opposite sides of a $yz$-plane at $x$ for the expression to be non-zero. This is the condition that the interaction is crossing the surface, and the force contribution is included. In this way, the MoP is most clearly related to the force over area definition of mechanical stress. The early work of Tsai[65] postulated this form of stress in a molecular system, but the work of Todd *et al.*[61] derived it from the statistical mechanics for the first time and provided a convenient form to use in molecular simulations. This was originally obtained through a Fourier transform of the original Irving-Kirkwood equations.[1,61] Working in Fourier space has the effect of averaging in the lateral directions and so the pressure is for an infinite plane. Regardless of the contour between two molecules, the contour must cross the plane if these two molecules are located on either side of the plane. The Fourier transform assumes an infinite periodic domain in $y - z$ and so avoids the need to choose an integral contour. However, we can prove that, due to the mechanical equilibrium, the ensemble average of the MoP form in Eq. (54) is equivalent to Eq. (13) which is derived directly from the contour form of the pressure tensor (see Appendix A1 for derivation). It is worth noting that the kinetic term in Eq. (54) still has the Dirac delta function, which requires some treatment before it can be used in a molecular simulation. Previous work writes the delta function as the sum of its roots,[146] which physically correspond to any crossings of the plane. However, a kinetic pressure form that is more consistent with the configurational term can be obtained by taking the integral of the kinetic pressure over a time interval from $t_1$ to $t_2$ and using a change of variables $dt = dx_i/\dot{x}_i$ to write,

$$\begin{aligned} \int_t^{t+\Delta t} \overset{\text{MoP}}{\mathbf{P}}{}_x^K(x)\, dt &= \frac{1}{S_x}\sum_{i=1}^{N}\int_{t_1}^{t_2}\frac{\mathbf{p}_i p_{ix}}{m_i}\delta\big(x - x_i(t)\big)dt \\ &= \frac{1}{S_x}\sum_{i=1}^{N}\int_{x_i(t_1)}^{x_i(t_2)}\frac{\mathbf{p}_i p_{ix}}{m_i \dot{x}_i}\delta\big(x - x_i(t)\big)\, dx_i \\ &= \frac{1}{S_x}\sum_{i=1}^{N}\mathbf{p}_i\big[H\big(x - x_i(t_1)\big) - H\big(x - x_i(t_2)\big)\big] \\ \overset{\text{MoP}}{\mathbf{P}}{}_x^K(x) &= \frac{1}{2\Delta t S_x}\sum_{i=1}^{N}\mathbf{p}_i[sgn(x - x_i(t_1)) - sgn(x - x_i(t_2))] \end{aligned} \tag{55}$$

The final line uses $H(x) = 1/2(sgn(x) + 1)$ to show the similarity of form to the common MoP configurational part and takes the average of the time integral on the left hand side $\int_t^{t+\Delta t} \overset{\text{MoP}}{\mathbf{P}}{}_x^K\, dt \approx \Delta t\, \overset{\text{MoP}}{\mathbf{P}}{}_x^K$. Finally, we note that, to be consistent, the momentum flux in these surface definitions $\mathbf{p}_i$ are expressed relative to a streaming velocity measured over the same surface, so if $dS_{ix} = [sgn(x - x_i(t_1)) - sgn(x - x_i(t_2))]$ then $\boldsymbol{u}(x,t) = \sum_{i=1}^{N} m_i \dot{\mathbf{r}}_i\, dS_{ix} / \sum_{i=1}^{N} m_i\, dS_{ix}$.

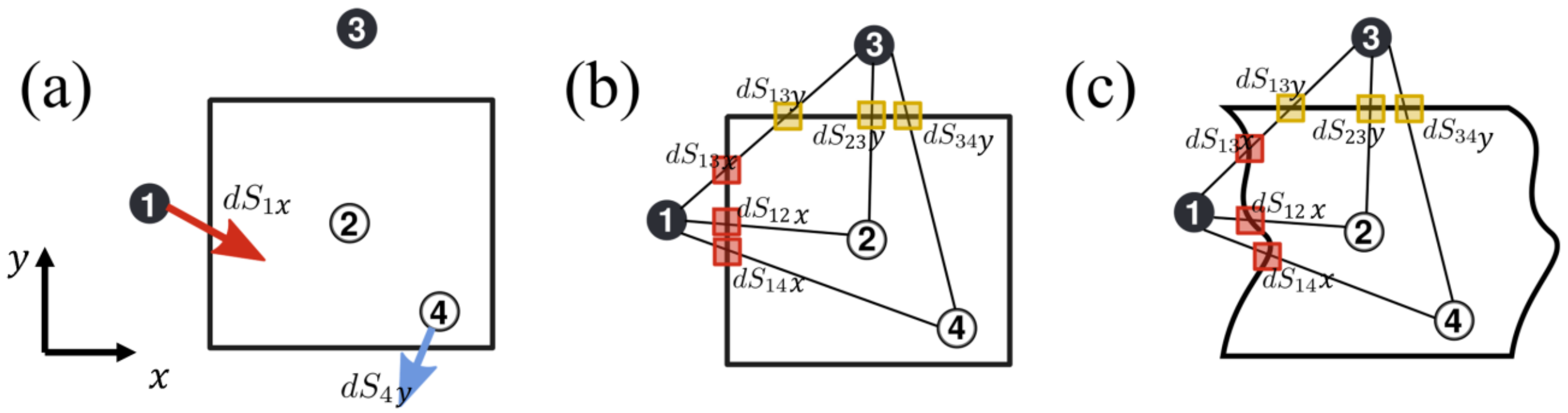

Figure 9. The local surface form of pressure, showing (a) the molecules 1 is entering the volume in $x$ and molecules 4 is leaving in $y$, both contributing to kinetic pressure; (b) the molecules interacting over the $x$-surface as red squares and $y$-surfaces as yellow squares contribute to the configurational pressure $\mathbf{P}_x$ and $\mathbf{P}_y$ respectively; and (c) a more general volume with an arbitrary surface requiring a ray-tracing style solution to identify crossings with additional terms for curvature and surface movement.[60] In both (b) and (c), the IK contour is used.

The original MoP formulation[61] only provides three pressure components as in Eq. (54), on an infinite plane. As the plane has a single normal component and is infinite in $y$ and $z$, this returns only a single pressure vector per plane. Han and Lee[147] used three mutually perpendicular planes converging at a point to obtain all nine components of the pressure tensor, and also limited the planes to a local region of interest. For example, using the boxcar function $\Lambda$ of Eq. (17), we can write this Local MoP (LP) for the kinetic and configurational pressure components on a local plane that is normal to the $y$-direction, *i.e.*, $P_{yx}, P_{yy}$ and $P_{yz}$ as,

$$\begin{aligned} \overset{\text{LP}}{\mathbf{P}}{}_y^K &= \frac{1}{2\Delta t \Delta S_y} \sum_{i=1}^{N} \mathbf{p}_i \overbrace{\left[sgn\big(y - y_i(t_1)\big) - sgn\big(y - y_i(t_2)\big)\right] \Lambda_x(x_i) \Lambda_z(z_i)}^{dS_{iy}} \\ \overset{\text{LP}}{\mathbf{P}}{}_y^C &= \frac{1}{4\Delta S_y} \sum_{i,j}^{N} \mathbf{F}_{ij} \underbrace{\left[sgn(y - y_i) - sgn\big(y - y_j\big)\right] \Lambda_x(x_k) \Lambda_z(z_k)}_{dS_{ijy}} \end{aligned} \tag{56}$$

where $\Delta t = t_2 - t_1$ and the function denoted as $dS_{iy}$ checks if molecule $i$ is crossing the plane (Figure 9a). Function $dS_{ijy}$ checks the crossing of the plane for intermolecular interaction path between $i$ and $j$ (Figure 9b). Here on the left-hand side of Eq. (56), we are using the shorthand notation for the average pressure integrated over a plane, $\overset{\text{LP}}{\mathbf{P}}_y \Delta S_y \approx \int_{S_y} \mathbf{P} \cdot d\mathbf{S}_y$. By localizing the pressure to a region of a plane, we are forced to choose a contour between molecules as different contours may or may not cross the corresponding plane sub-section. Using the IK contour, the boxcar function $\Lambda_x(x_k)$ in Eq. (56) identifies if a crossing $x_k = x_i + \lambda_k x_{ij}$ is on the surface, with $\lambda_k$ being the value of $\lambda$ along the line of contour integration at the point of crossing of the plane/surface. It is interesting to note that the kinetic pressure in this form is also now dependent on the trajectory "contour" of the molecules as they evolve in time, so that different integration methods or timesteps could result in different measured crossings and therefore a non-uniqueness in the kinetic pressure. In practice, molecules move by very small amounts in a time step, so any difference is unlikely to be apparent.

For now, we have derived the surface pressure forms of Eq. (54) (MoP) and Eq. (56) (LP) by taking the zero volume limit of the VA form of Eq. (52). However, the localized surface pressure is more rigorously derived by taking the derivative of the control volume pressure in Eq. (51),[148] which gives the six local surfaces pressures bounding an enclosed region,

$$\begin{aligned} \boldsymbol{\nabla} \cdot \int_V \left[\mathbf{P}^K(\mathbf{r},t) + \mathbf{P}^C(\mathbf{r},t)\right] dV \\ &= \left(\overset{\text{LP}}{\mathbf{P}}{}_x^+ - \overset{\text{LP}}{\mathbf{P}}{}_x^-\right)\Delta S_x + \left(\overset{\text{LP}}{\mathbf{P}}{}_y^+ - \overset{\text{LP}}{\mathbf{P}}{}_y^-\right)\Delta S_y + \left(\overset{\text{LP}}{\mathbf{P}}{}_z^+ - \overset{\text{LP}}{\mathbf{P}}{}_z^-\right)\Delta S_z \\ &= \oint_S \left[\mathbf{P}^K(\mathbf{r},t) + \mathbf{P}^C(\mathbf{r},t)\right] \cdot d\mathbf{S} = \frac{d}{dt}\int_V \rho \boldsymbol{u} dV \end{aligned} \tag{57}$$

where $\overset{\text{LP}}{\mathbf{P}}{}_\alpha^+$ and $\overset{\text{LP}}{\mathbf{P}}{}_\alpha^-$ $(\alpha = x, y, z)$ denote the LP pressure tensor on upper (+) and lower (-) $\alpha$-surfaces, corresponding to front/back, right/left, and top/bottom pairs in the *x*-, *y*- and *z*-directions, respectively (see Figure 4a for the coordinate system). The pressure on all six surfaces of an enclosed volume is exactly equal to the momentum change in time, (to machine precision) due to the conservative property of the finite volume form.[115] This is denoted by the final line in Eq. (57) which states that the integral around the bounding surface is equal to the change inside that volume. It is also possible to use this technique for any volume, for example the surface pressure for spherical volumes have been derived in the literature[144] and we consider more general surfaces in Section 4.5). To show this conservation, we introduce the shorthand $P^{\pm} = P^+ - P^-$, and define Advection to include the convective term and kinetic pressure, a Forcing term including configurational pressure on a surface, and plot both against the change inside the volume, called Accumulation in Figure 10.

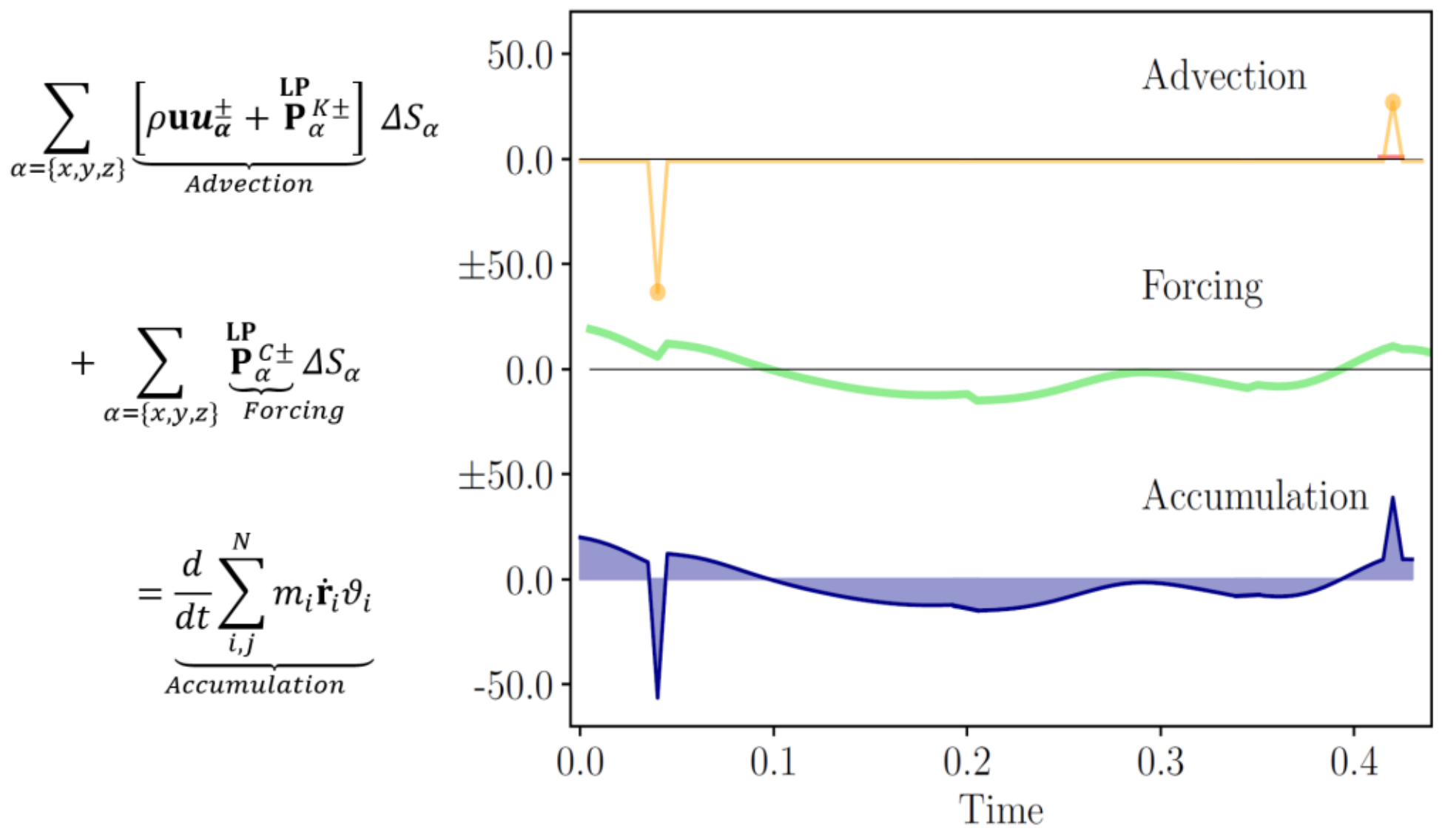

Figure 10. Conservation of momentum for a control volume in a molecular dynamics simulation, where the sum of the molecules crossing the surface (Advection) and forces acting over the surface (Forcing) is equal to momentum change inside the volume (Accumulation).

This process of surface flux derivation can also allow more general volumes providing a form of pressure on any arbitrary surface (*e.g.*, a rippling surface), as shown in the next section. Mechanically, each of the terms can be viewed as the tractions in a Cauchy tetrahedron, so using the three orthogonal planes on the top surfaces with a traction force vector on each, *c.f.* Figure 4a, the three surface traction vector can be assembled to give $\mathbf{P}(\mathbf{r},t) = \left[\overset{\text{LP}}{\mathbf{P}}{}_x^+, \overset{\text{LP}}{\mathbf{P}}{}_y^+, \overset{\text{LP}}{\mathbf{P}}{}_z^+\right]^T$ which is a nine-component pressure tensor. A similar tensor could be defined for the bottom set of surfaces defining the other tetrahedron that together makes up the cube volume. It is worth noting that these surface pressures can be derived directly as surface localization of phase space quantities[148] that satisfy the requirements of statistical mechanics[78]. However, the real strength in these instantaneous stresses on a bounding surface is that we can get the stress at any instant which is directly responsible for a corresponding momentum change inside the volume.

## 4.5 A moving reference frame

For interactions crossing an arbitrary surface which is itself changing in time, we can define a volume which follows a feature of the simulation. For example, we consider a two-dimensional example of a liquid-vapor interface (projected onto the $yz$-plane) defined by a generalized function $\xi(y,z,t)$. Using this moving interface function, the volume average equation of Eq. (51) can be repurposed with the boxcar function redefined as $\Lambda_x = H(x_i - x + \Delta x/2 + \xi(y,z,t)) - H(x_i - x - \Delta x/2 + \xi(y,z,t))$. Taking the derivative of Eq. (51) gives the surface pressure, as before, but it is now on a curved and moving surface and we refer to it as the surface flux (SF) pressure $\overset{SF}{\mathbf{P}}$. Working through the mathematics, the kinetic and configurational SF pressure tensor can be expressed as follows,[149]

$$\begin{aligned}
&\rho \boldsymbol{u} u_x + \overset{\text{SF}}{\mathbf{P}}{}_x^{K+} \\
&= \frac{1}{\Delta t \Delta S_x} \sum_{i=1}^{N} m_i \dot{\mathbf{r}}_i \int_{t_1}^{t_2} \left[ \dot{x}_i + \overbrace{\dot{y}_i \frac{\partial \xi_i^+}{\partial y_i} + \dot{z}_i \frac{\partial \xi_i^+}{\partial z_i}}^{Kinetic\ Curvature} + \overbrace{\frac{\partial \xi_i^+}{\partial t}}^{Surface\ Evolution} \right] dS_{ix}^+ dt \\
&\overset{\text{SF}}{\mathbf{P}}{}_x^{C+} = \frac{1}{2\Delta S_x} \sum_{i,j}^{N} \mathbf{F}_{ij} \int_0^1 \left[ x_{ij} + \underbrace{y_{ij} \frac{\partial \xi_\lambda^+}{\partial y_\lambda} + z_{ij} \frac{\partial \xi_\lambda^+}{\partial z_\lambda}}_{Configurational\ Curvature} \right] dS_{\lambda x}^+ d\lambda
\end{aligned} \tag{58}$$

Eq. (58) is identical to the LP pressure of Eq. (56) but with all the extra terms accounting for interface curvature and movement. The surface values at the point that the molecular trajectory or intermolecular interaction cross are denoted by $\xi_i^+$ and $\xi_\lambda^+$, respectively, while $dS_{ix}^+$ *and* $dS_{\lambda x}^+$ are functions which are non-zero only if the crossings occur on the particular patch of surface (a generalization of the signum functions of Eq. (56)). The underbraces highlight the physical meaning of the various terms. The Surface Evolution term accounts for the movement of the interface in time, and the Kinetic and Configurational Curvature terms ensure that $y$ and $z$ components are included on the $x$-surface. The convective term is included on the left-hand side as surface movement makes it more difficult to determine what is convection and what is kinetic pressure. By introducing the definition for the surface normal vector $\tilde{\mathbf{n}}_x = \boldsymbol{\nabla}_\alpha(\xi - x_\alpha)/|\boldsymbol{\nabla}_\alpha(\xi - x_\alpha)|$ and evaluating the integrals in Eq. (58), we obtain a form of pressure purely in terms of the surface normal (for a full derivation please see Ref. [60]). We note that the kinetic term in Eq. (58) has been written as a time integral, which provides the symmetry with the configurational term discussed in relation to Eq. (56) and makes the (numerical) implementation identical to the configurational part. By taking the integral in Eq. (58), we arrive at

$$\begin{aligned} \rho \boldsymbol{u} u_x + \overset{\text{SF}}{\mathbf{P}}{}_x^{K+} &= \frac{1}{\Delta t \Delta S_x} \sum_{i=1}^{N} m_i \dot{\mathbf{r}}_i \frac{\mathbf{r}_{i12} \cdot \tilde{\mathbf{n}}_x}{|\mathbf{r}_{i12} \cdot \tilde{\mathbf{n}}_x|} dS^+ + \frac{1}{\Delta S_x} \sum_{i=1}^{N} m_i \dot{\mathbf{r}}_i \vartheta_t \\ \overset{\text{SF}}{\mathbf{P}}{}_x^{C+} &= \frac{1}{2\Delta S_x} \sum_{i,j}^{N} \mathbf{F}_{ij} \frac{\mathbf{r}_{ij} \cdot \tilde{\mathbf{n}}_x}{|\mathbf{r}_{ij} \cdot \tilde{\mathbf{n}}_x|} dS^+ \end{aligned} \tag{59}$$

The Surface Evolution term is contained in the function $\vartheta_t = \left[H\big(\xi(t_2) - x_i(t_2)\big) - H\big(\xi(t_1) - x_i(t_2)\big)\right]\Lambda_y(y_i(t_2))\Lambda_z(z_i(t_2))$, where molecule positions are fixed and we count how many molecules have left or entered the volume due to the movement of the surface in the time interval between $t_1$ and $t_2$. For the kinetic term $\mathbf{r}_{i12} = \mathbf{r}_{i2} - \mathbf{r}_{i1}$ is the line of time evolution of a molecule $i$ between $t_1$ and $t_2$ which mirrors the configurational term's intermolecular (IK) contour $\mathbf{r}_{ij} = \mathbf{r}_j - \mathbf{r}_i$. Assuming the IK contour, the equation for a line is $\mathbf{r}_\lambda = \mathbf{r}_{i1} + \lambda \mathbf{r}_{i12}$ where the value of $\lambda$ at the point of crossing of the surface is $\lambda_k$. There is no closed form equation to get $\lambda_k$ in general, but we can triangulate or split the surface into patches and use a ray-tracing process to obtain the point of intersection of a line and the surface.[150] Once we obtain this crossing, it can be inserted in the following expression for the use of Eq. (59) in molecular simulations,

$$dS^+ = \sum_{k=1}^{N_{roots}} [H(1 - \lambda_k) - H(-\lambda_k)]\, \Lambda_y(y_k)\Lambda_z(z_k) \tag{60}$$

which is non-zero only if the line crossing the surface is between the start $\lambda_k = 0$ and finish $\lambda_k = 1$ of the line, and the point of crossing in the $y$- and $z$-direction, $y_k$ and $z_k$, respectively, fall between the surface patch limits. Figure 9c illustrates the use of Eq. (59) on a general surface.

If we neglect the Surface Evolution term in Eq. (58), the general form of surface pressure is similar to the spherical and cylindrical pressure tensor presented in Eqs. (20), (22), (28), and (30). The surface normal $\tilde{\mathbf{n}}_x$ in Eq. (59) behaves like the radial ($\hat{\mathbf{R}}$) or azimuthal ($\hat{\boldsymbol{\phi}}$) unit vectors, providing the pressure tensor which is aligned to the normal or tangential vector for a general surface which varies in both $y$ and $z$, *i.e.*, $\tilde{\mathbf{n}}_x = \tilde{\mathbf{n}}_x(y, z)$. The function $H(1 - \lambda_k) - H(-\lambda_k)$ in Eq. (60) collects only the interactions crossing the surface, as the function $H(\lambda_k)H(1 - \lambda_k)$ does in Eqs. (20), (22), (28), and (30). The extra localization of the $\Lambda$ functions could have been included in the cylindrical and spherical forms of pressure if inhomogeneity along the surface were of interest. The general form presented in Eq. (59) can provide a detailed picture of molecular stacking and its effect on pressure near a much more complex surface, for example the intrinsic interface, obtained by refitting each time to a set of molecules where a liquid meets a vapor.[149] This uses sine and cosine functions with wavelengths chosen to allow fitting down to the molecular spacing. To capture instantaneous fluctuations about the spherical shape, spherical harmonics or similar could be used to capture the details of the molecular stress structure. These approaches quickly become cumbersome in general, so in the next section, we present a more pragmatic approach.

## 4.6 Coordinate transforms

Evaluating the pressure form in Eq. (59) on a time-evolving surface requires an interface definition, together with an interaction calculation for every pair of molecules crossing that interface. As MD becomes a more common simulation tool, researchers are increasingly tackling more complex interface geometries, which require a more general fitted surface. This makes interface tracking increasingly more complex and the resulting averaging grid can become impossibly deformed. Instead, in this section we discuss a process of collecting pressure values on a uniform grid and performing a transform afterwards. Transforming the pressure requires two steps. The first is a rotation of the pressure tensor so the pressure is aligned with the normal to the surface. The second is a mapping so the pressure we collect can be obtained either at a distance from the surface or as a function moving along that interface. Through this process, we also highlight an important subtlety in pressure tensor studies, that the correct pressure tensor for a problem is dependent on measuring location and alignment.

As an example, we consider the NEMD boiling simulation shown in Figure 11, a case of great practical interest.[151–153] Here a solid wall of tethered molecules is heated by a thermostat from the bottom and a phase change occurs in the liquid, starting at the bottom of a nano-scale square pore in the wall. The liquid is set up as a finite film with a large gas region above, and a bubble is allowed to nucleate and grow. The simulation is pseudo-two-dimensional with a nominal thickness in $z$, and with periodic boundaries. The bubble grows initially inside the square pore, before extending beyond and forming a roughly circular shape. Figure 11a shows the density field where the yellow region is liquid with density of 0.7 in LJ units and the dark blue region is vapor with density of 0.05 in LJ units. There is no unique way to identify the liquid-vapor interface. In this work, a simple thresholding operation $\rho > 0.3$ (LJ units) is performed to identify the liquid, followed by a gradient operation to get the interface location where the density gradient is non-zero. These interface pixels are fitted using a least-square algorithm, with a circular arc to the part of the bubble above the wall.  The fitted arc has a radius and angle used to rotate the pressure field $P_{xx}$ in Figure 11b:[154]

$$\begin{aligned} P_{RR} &= P_{xx}\cos^2\theta + P_{yy}\sin^2\theta + P_{xy}\sin(2\theta) \\ P_{\theta\theta} &= P_{xx}\sin^2\theta + P_{yy}\cos^2\theta - P_{xy}\sin(2\theta) \end{aligned} \tag{61}$$

It is worth noting that the pressure tensor is a measure of the alignment of force with a given coordinate axis, so any rotation of the tensor is equally valid. Upon transform, this exposes a clear polar (tangential) pressure $P_{\theta\theta}$ around the interface in Figure 11c, similar to a hoop stress in a solid pressure vessel, which is what holds the bubble's shape. It is this tangential contribution that will be significant in a Kirkwood-Buff surface tension calculation. In order to get $P_{\theta\theta}$ at a given radius, we need to average all values at the same radius over $\theta$. Here we apply this on the top half of the bubble, to avoid the near-wall region. The integration limits are shown in Figure 11c by the double ended arrow between the angle at the bottom denoted by $\theta_b$ and the angle at the top denoted by $\theta_t$. We use a mapping or projection from the Cartesian grid to a polar grid. Interestingly, instantaneous local fluctuations about the perfect circle can be seen in this projection in the blue line of tangential pressure in Figure 11d. We average along the interface, in this case between $\theta_b = -90°$ and $\theta_t = 90°$ for both radial and tangential pressure to obtain the plot in Figure 11e. This shows both the imbalance of radial pressure $P_{RR}$, which is driving the interface to grow, and the contribution to surface tension due to the tangential pressure $P_{\theta\theta}$. The resulting pressure is similar to the one that would be obtained using Eqs. (20) and (22), but we have obtained it using data collected from the simulation on a uniform grid and by fitting the interface afterwards. This technique is ideal for existing molecular dynamics software packages which do not have a rich selection of pressure calculation methods inbuilt, as discussed in Section 5.3, allowing uniform fields to be repurposed.

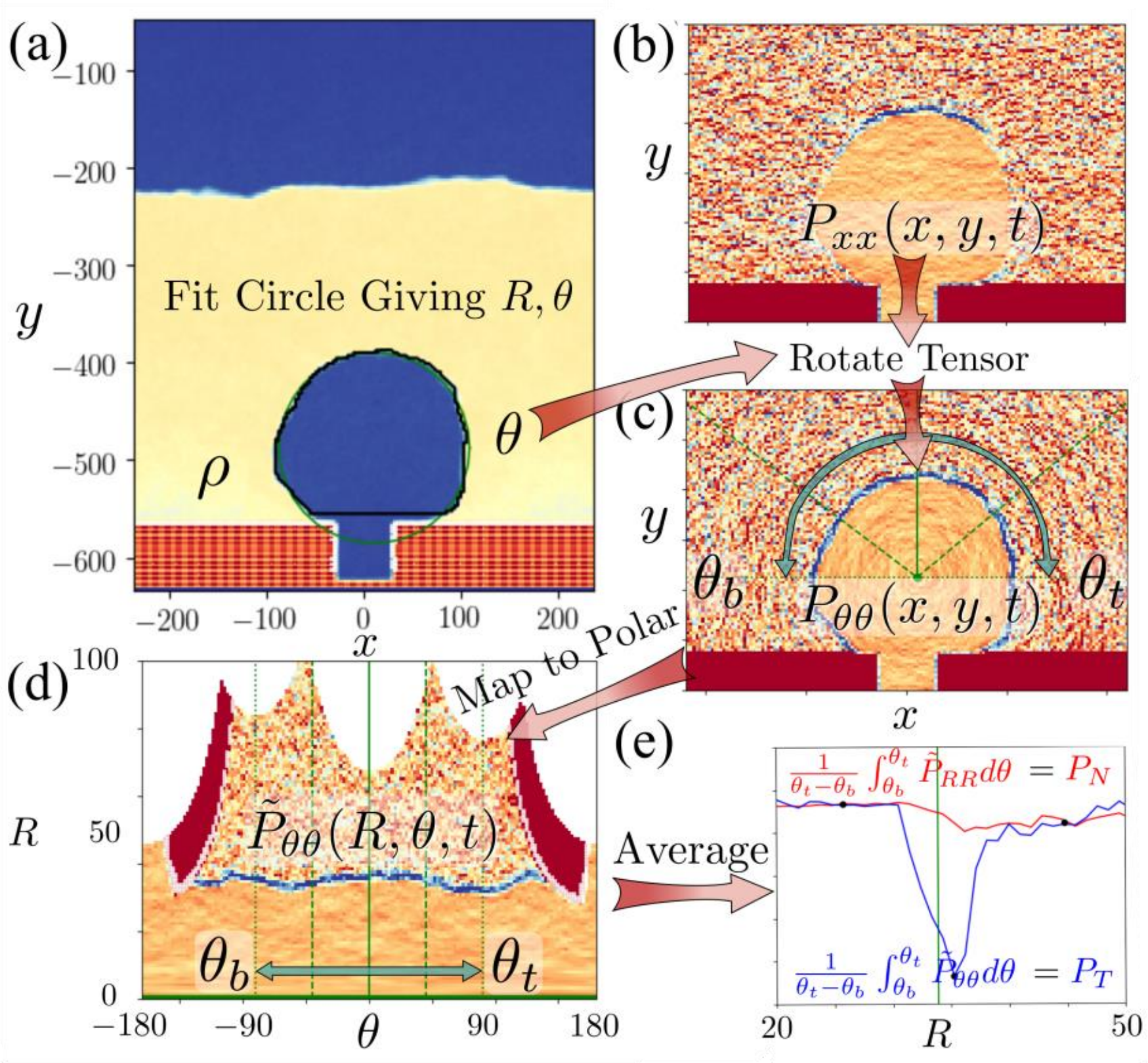

Figure 11. An example of the use of mapping in obtaining a pressure relevant to the geometry of a problem. The example is a single snapshot in time of a NEMD boiling simulation, where a bubble starts from a square notch on a wall and expands into a circular bubble shown here. The bubble is clear in the density field of (a) with blue vapor density and yellow liquid density. The identified interface points are shown as black pixels and the green line shows a fitted circle arc used to get radius and angle. (b) The pressure field of $P_{xx}$

component with blue indicating large negative values. (c) The pressure rotated using the angle from the fitted circle and Eq. (61) to give tangential pressure $P_{\theta\theta}$, with green lines shown at $\theta_b = -90°$, -45°, 0°, 45° and $\theta_t = 90°$ to guide the eye. The arrow shows the whole 180° arc. (d) The mapped field using a cartesian to polar mapping, where the green lines correspond to the ones from (c) between $\theta_b$ and $\theta_t$. (e) The average normal and tangential pressure over the 180 arc.

## 4.7 Statistical uncertainty of different pressure methods

One of the main difficulties of pressure tensor calculations is the high level of noise relative to quantities such as $\rho, \boldsymbol{u}$ and $T$. This is clearly observed in Figure 11b where a noisy pressure field gives a grainy appearance, which is absent from the density field in Figure 11a. It is worth observing that what we term "noise" here in NEMD are more concretely fluctuations about the time-evolving averaged quantities of interest, which tends to obscure them. In equilibrium systems, this noise can actually be the quantity of interest, such as diffusion or viscosity obtained from the Green-Kubo formula,[43,44] or certain thermodynamic functions (*e.g.*, heat capacity) that are a measure of fluctuations.[155] As discussed in Section 4.3, these fluctuations might contribute to the turbulent eddies, and so they are important for an overall understanding of the flow. Hadjiconstantinou *et al*[156] estimated that collecting pressure is orders of magnitude worse in terms of statistics, which, they argue, makes coupling, passing averaged MD pressure values to be used as boundary conditions in a continuum solver, untenable. For NEMD simulations, getting good statistics becomes increasingly problematic as time-evolving events can depend on a chaotic trajectory, making it difficult to create a consistent ensemble. The example of boiling in Section 4.6 makes this clear, where nucleation occur at different times in the members of an ensemble, and the bubble growth proceeds in a varied and apparently stochastic way. A general discussion on the topic of pressure noise is difficult compared to other microscopic properties,[157] as the statistical requirements are case specific, depending on rate of time evolution in the system and choice of averaging volume size. Often, we are forced to use a coarse spatial resolution in order to provide sufficiently well-behaved pressure measurement.

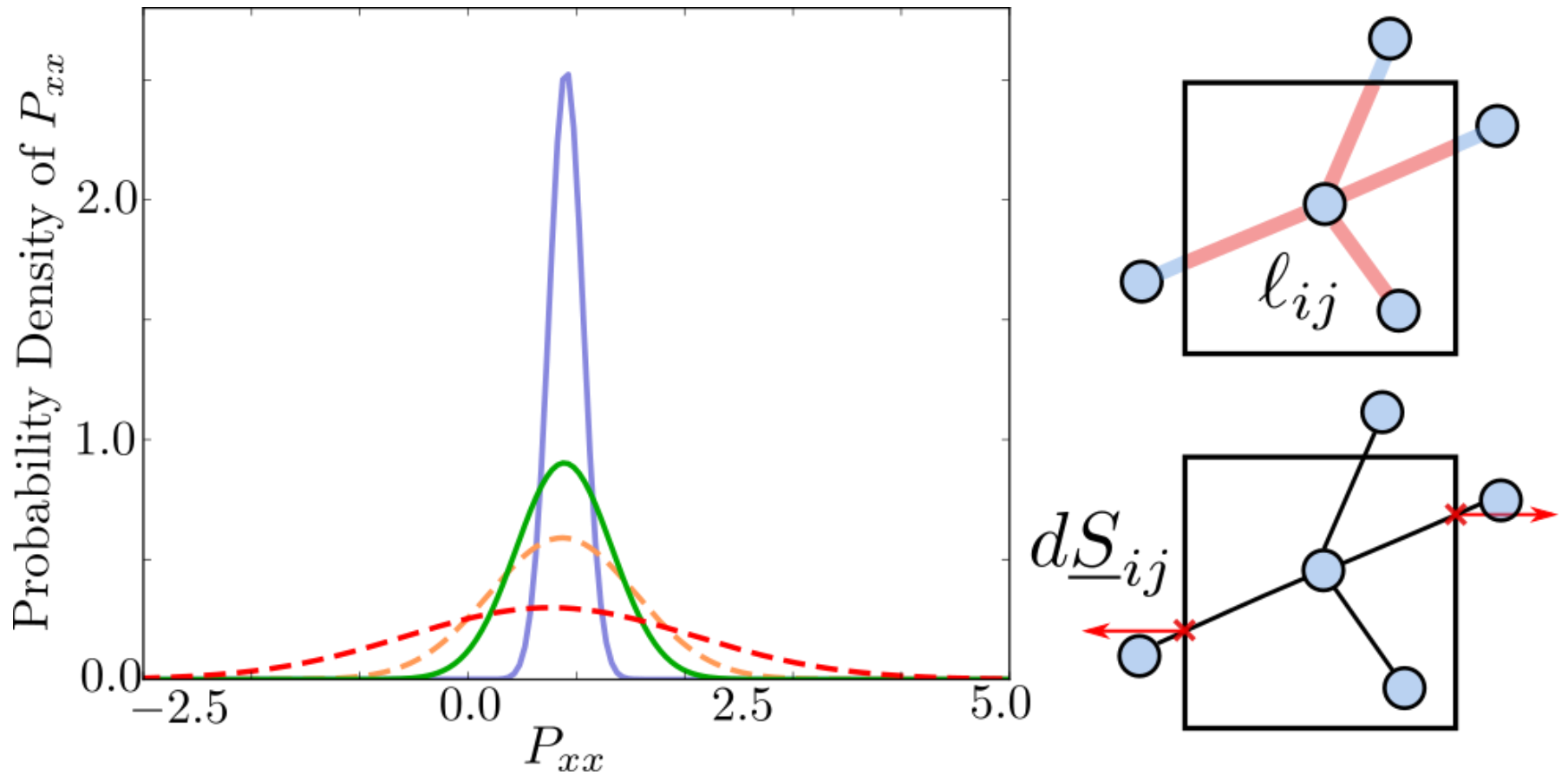

Figure 12. Comparison of the probability distribution function of $P_{xx}$ using the VA form ($\ell_{ij}$ is length of line in a volume shown schematically in red on the top right) and LP form ($d\underline{S}_{ij}$ is surface crossing shown

schematically as red crosses with surface normal on the bottom right). For a reduced cell side length of $13.7\sigma$, the blue solid line on the main plot is the VA pressure, and the dotted yellow line is the LP pressure. For a reduced cell side length of 6.8 $\sigma$, the green solid line is the corresponding VA pressure, and the dotted red line is the LP pressure. Data are taken from Ref. [158].

To get a sense of the magnitude of this noise, Figure 12 shows the distribution of measured pressure in a sub-volume of a large periodic simulation box of reduced density 0.8.[158] The distributions are Gaussians for boxes of this size. Two cubic volume sizes are considered, the large volume has a reduced cell side length of $L = 13.7\sigma$ which will contain about $N = 2000$ molecules, while the smaller volume has a reduced side length of $6.8\sigma$ with around $N = 250$. The VA pressure using Eq. (52) and the LP pressure using Eq. (56) are shown by solid and dotted lines, respectively. The average pressure is roughly the same for all cases with value $\langle P_{xx} \rangle \approx 0.9$ reduced units, shown by the similar peak locations of the Gaussians in Figure 12. Interestingly, the VA measures show a much lower spread in the distribution, *e.g.* standard deviations for the large volume $\text{std}\left(\overset{\text{VA}}{P_{xx}}\right) = 0.16$ versus $\text{std}\left(\overset{\text{LP}}{P_{xx}}\right) = 0.68$, both in LJ units. This means standard deviation is more than 4 times higher in the surface pressure measurement than the volume average for the same number of samples. For the smaller volume case, where surface to volume ratio is improved by a factor of two in favor of the surface measurement, the standard deviation of the LP pressure is still about 3 times that of the volume average. One reason for this is apparent from the way pressures are obtained, as shown schematically in Figure 12 (top right and bottom right). The VA scheme uses a continuous fraction of all interactions based on the length of line in the volume, which provides some smoothing as well as averaging from all interactions. Meanwhile, the LP pressure includes a contribution only if the surface is crossed, so fewer interactions are counted and these change abruptly as molecules move until interactions suddenly no longer cross a surface. Also, for $\overset{\text{LP}}{P_{xx}}$ only a single surface is used compared to the entire volume for the VA, further reducing the samples obtained in practice.

In conclusion, the VA pressure performs better at reducing noise, giving between 3 to 4 times lower standard deviation than a surface definition in this example. However, the conservative properties of the surface definition, as well as the ability to track complex geometries, makes surface pressure preferable in some cases.

# 5. Challenges and future directions

## 5.1 Controversies over the microscopic pressure tensor

The microscopic pressure or stress tensor has been a controversial topic since the 1950s, largely due to the arbitrary contour involved in the formalism (Eq. (9)). In fact, even the concept of the virial pressure introduced in 1870 by Clausius[81] was not free of controversy. The form of the virial pressure was questioned in 1895, by a Colonel Basevi,[159] who argued that by performing the integration in time that the kinetic and configurational parts should cancel. This argument was refuted by at least two articles[160,161] with A. Gray[161] noting that "*Colonel Basevi has, it seems to* *me, overlooked the fact that in the theorem it is the forces acting on each particle relatively to the assumed axes, and the corresponding motions that must be taken into account*". Such confusions about the assumed axis and corresponding motion are apparently still a subject of confusion today, together with the non-uniqueness of the pressure tensor itself.

The non-uniqueness problem of the microscopic pressure tensor is well-known in the statistical mechanical community. It was implicit in the paper of Kirkwood and Buff[70] in 1949, and the arbitrariness of the force acting across a surface element was then discussed in an Appendix of the seminal paper by Irving and Kirkwood.[1] Perhaps due to the introduction of a surface, Irving and Kirkwood's warning did not attract much attention, although it was apparently noticed by Harasima and others[67,162] when they studied the surface tension of liquids. The non-unique nature of the microscopic pressure tensor became a focus in the 1980s, a time when molecular simulation was emerging as a technique to study complex systems that are critical in engineering, biology, and physics. Schofield and Henderson,[2] for the first time, crystalized the ambiguity in the microscopic pressure tensor as an emergent property of the arbitrary contour shown in Eq. (9). Since then, many attempts have been made to find reasonable arguments and additional constraints[78,163–168] which limit the choice of the contour. Until now, no consensus has been reached in the field,[39,169] and there is no convincing justification for choosing one contour definition over the other for general cases. The non-uniqueness of the microscopic pressure tensor reflects the fact that there is no unequivocal way to assign a force (mechanical route, Eq. (9)) or potential energy (thermodynamic route, Eq. (37)) to a point $\mathbf{r}$ in space. In practice, the pressure is not measured at a point, but over a volume or surface, so the functional form of this kernel also matters, together with its location and shape. Any rotation or coordinate transform of a stress tensor changes the relative magnitude of the tensor components according to the orientation of the measurement. This has led to the concept of principal components in stress analysis,[170] *i.e.* a rotation of the tensor, so that shear components are zero, providing an invariant or unique stress. From a fluid dynamical perspective, the local pressure tensor is subject to the so-called "gauge transform" where one can add a constant, or even the curl of any vector field, to the momentum density without affecting the system dynamics.[2,61] This is because it is the gradient of the stress tensor that is well-defined, and not the stress tensor itself.

The other noteworthy controversy over the definition of the stress tensor was raised by Zhou in 2003,[171] where the inclusion of the kinetic term in the stress definition was questioned. This controversy is, in part, due to the differing definitions of pressure/stress tensor in the solid mechanics, thermodynamics and fluid mechanics literature. In the solid mechanics literature, the Cauchy stress tensor is defined in terms of forces at zero temperature (*i.e.,* no kinetic part). Often the temperature dependence is included using an extra term in the continuum. Zhou's argument, however, has been refuted by multiple studies. Admal and Tadmor[78] showed that Zhou's conclusions result from not considering the difference between absolute and relative velocities. They also demonstrated that the kinetic contribution to the stress is significant even for solid systems at a finite temperature. Subramaniyan and Sun[172] showed the importance of temperature on stress in a thermo-elastic study using molecular dynamics. Hoover *et al*.[173] achieved an excellent agreement between atomistic mechanics and continuum mechanics provided that both kinetic and configurational contributions to the stress tensor are considered. It has been shown that the kinetic contribution to the stress is a direct consequence of the canonical transformation.[8,174] We also note that the kinetic term is essential in the pressure and stress tensor definitions for thermodynamic consistency in the ideal gas limit (Section 3). Away from thermodynamic equilibrium, kinetic pressure is defined in terms of peculiar velocity, the molecular velocity left after subtracting the streaming velocity of the flow. This contribution is essential in the pressure, gradients of which can drive flow in fluid dynamics, as well as in the shear stress between the molecules which underpins fluid viscosity. The importance of kinetic pressure is most apparent when molecular simulation includes turbulent flow (Section 4.3), where the kinetic contribution is shown in Figure 7b to be as large as the configurational shear stress in turbulent flow and essential as a direct continuation of Reynolds stress below the scale of the measuring grid.

As molecular simulations find wider use both in fundamental science and, increasingly, in industrial applications, the necessity of finding an agreed definition of the microscopic pressure/stress tensor has become more critical than ever. We now consider a few promising contributions in this direction.

Motivated by the thermodynamic concept of pressure that is conjugate to a finite volume instead of to a point, Shi et al.[35] showed that by spatially averaging the local (non-unique) pressure tensor over a small region of space of molecular dimensions, it is possible to define a coarse-grained (CG) microscopic pressure tensor that is unique, and free from ambiguities in the definition of the local pressure tensor. In the case of fluids confined in a slit-shaped pore with $z$-axis perpendicular to the flat surface, such unique CG pressure tensor in $k^{\text{th}}$ bin ($k = 1, 2, 3, \ldots$) along the $z$-axis is given by:

$$\begin{aligned} \overset{\text{CG}}{\mathbf{P}}_k &= \frac{1}{\Delta \mathbf{r}_k} \int_{\Delta \mathbf{r}_k} \mathbf{P}(\mathbf{r})\, \mathrm{d}\mathbf{r} \\ &= \frac{1}{\Delta z_k} \int_{\Delta z_k} \mathbf{P}(z)\, \mathrm{d}z \end{aligned} \tag{62}$$

where the local (averaging) volume of the $k^{\text{th}}$ bin is $\Delta \mathbf{r}_k = L_x L_y \Delta z_k$, $L_x$ and $L_y$ are the constant lateral dimension of the pore surface in the $x$- and $y$-directions, respectively, and $\Delta z_k$ is the characteristic length (width) of the $k^{\text{th}}$ bin that leads to a unique CG pressure tensor. This CG pressure tensor has the same appearance as the conventional VA definitions[110,143,175,176] in Eqs. (51)-(52). However, the CG pressure is essentially different from the VA pressure in terms of the choice of the averaging region. For the CG pressure tensor, the averaging volume is constrained to give a unique CG pressure, while for the VA pressure tensor, the averaging volume is unrestricted and thus the resulting VA pressure can still be subject to the arbitrary choice of the integral contour.

To find the proper averaging region that will lead to a unique CG pressure, Shi et al.[35] carried out the integration of the local tangential pressure over the $z$-direction analytically. They found that the contour path connecting particles $i$ and $j$, upon integration, is fully dictated by a function $f_C(\lambda_{ij})$, where $\lambda_{ij}$ is the linearly scaled $z$-distance from particle $i$ to particle $j$ (assuming $z_i < z_j$); thus $\lambda_{ij} = 0$ amounts to $z = z_i$ and $\lambda_{ij} = 1$ is $z = z_j$. Taking advantage of the symmetry of the contour path due to the indistinguishability of particles, 10 contour definitions were designed, equivalent to 10 unique functional forms for $f_C(\lambda_{ij})$, as shown in Figure 13a. These 10 contours include IK, H, IK-VR, and H-VR definitions that are introduced in Figure 3. Using these 10 types of contour definitions, they found that integrating the non-unique local tangential pressure over a certain $z$-distance leads to convergent integral results (within numerical uncertainty) that are independent of the arbitrary contour definitions (Figure 13b).[35] The CG pressure tensor that is defined between these convergence points (see characteristic length $\Delta z_1$, $\Delta z_2$ marked in Figure 13b for example) appears to be unique. Because these characteristic lengths are comparable to the thickness of the adsorbed layer (see density profile in Figure 13b), the CG pressure tensor has direct physical significance, representing the microscopic pressure in an adsorbed layer.

The proposed CG scheme may serve as a unified solution towards a unique microscopic pressure tensor that is free from ambiguities in contour definitions, averaging volume and shape, and measurement locations. Future studies should focus on providing further simulation evidence, and ideally, rigorous mathematical proof to support the existence of this unique CG pressure tensor. Systems that are of particular interest for testing include those having curved interfaces and arbitrary geometries, and those with moving reference frames.

Another possible aid to the problem of non-uniqueness is the conservative properties of the control volume form of pressure outlined in Section 4.4 and demonstrated in Figure 10. This simply states that the pressure measured on surfaces which form an enclosed volume must exactly equal the momentum change inside.[148,177] This is valid arbitrarily far from equilibrium and can be checked for any form of interaction contour, and with deforming or moving volumes,[60] ensuring the pressure measurements satisfy Newton's law. A different choice of volume or contour will simply redistribute the contribution to different terms in the tensor (as contributions are counted on different faces, for example). This exact equality between pressure and momentum change can help restrict the definition of the microscopic pressure tensor to the one that ensures that we do not violate Newton's law.

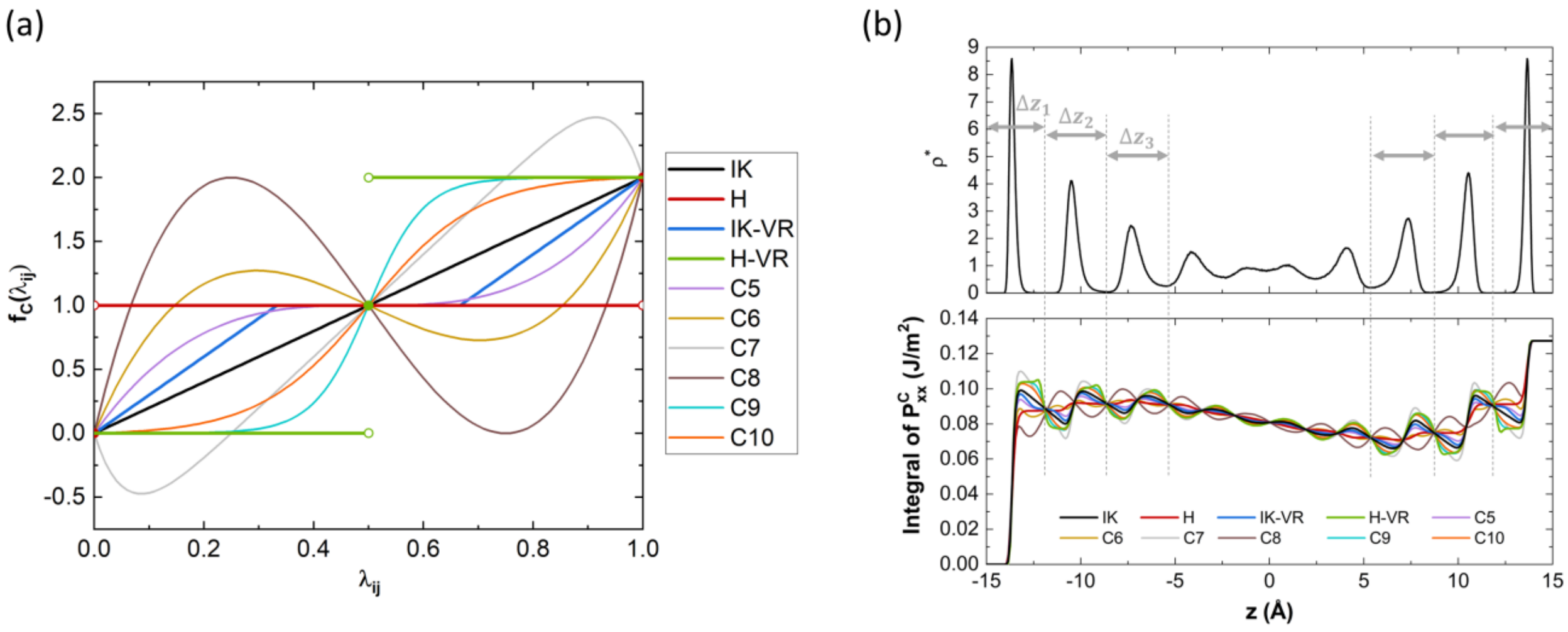

Figure 13. Simulation evidence for the uniqueness of the CG pressure tensor defined in Eq. (62).[35] (a) Graphs showing 10 functional forms for $f_C(\lambda_{ij})$ with $\lambda_{ij} = (z - \min(z_i, z_j))/|z_{ij}|$, corresponding to 10 unique contour definitions. (b) Reduced number density profile (upper panel) for LJ argon adsorbed in a structureless carbon slit pore and the integral of the (configurational) local tangential pressure $P^C_{xx}(z)$ over $z$-direction using the 10 contour definitions (bottom panel). Only the adsorbate-adsorbate interactions contribute to the tangential pressure calculations. In this case, the characteristic length $\Delta z_k$ that can lead to a unique CG pressure tensor were chosen to be those that are comparable to the thickness of the adsorbed layer, and they are marked in the plot as $\Delta z_1$, $\Delta z_2$, *etc.* Adapted from Ref. [35], with the permission of AIP Publishing.

## 5.2 Complex systems interacting with many-body and long-range potentials

So far, we have only focused on the pressure tensor for systems of discrete particles that interact with short-range pairwise potentials. It is of practical interest to extend these formalisms to molecular and material systems that are controlled by more realistic interaction potentials. For example, in biological systems, the internal structure of the molecule is mainly dictated by many-body intramolecular interactions such as angular, torsion and improper potentials. As for intermolecular potentials, in addition to the short-range dispersion interactions, long-range Coulombic interactions are commonly present in the system due to the uneven distribution of charges in the molecule.

In general, the pressure tensor can be expressed in two equivalent forms: an *atomic* form and a *molecular* form. The *atomic* pressure tensor is defined in such a way that all forces on each atom should be evaluated explicitly;[178–180] these include contributions from intermolecular interactions and intramolecular interactions (*e.g.*, bonded and bending forces, and constraint forces imposed by the SHAKE algorithm[181]). The *molecular* pressure tensor takes the molecule to be a rigid body (*i.e.*, rigid body approximation[182]), and only (non-bonded) intermolecular interactions contribute to the pressure tensor.[74,83,178–180] Compared to the atomic pressure tensor, this molecular formalism neglects the intramolecular interactions, and thus loses the mechanical details of the internal structure of the molecule. In addition, the implementation of the molecular pressure tensor requires defining the center of mass (COM) at which to localize the momentum of a molecule.[180] This definition of COM is straightforward for small molecules, but is by no means intuitive for solid materials or polymer chains, as they are usually modeled as infinitely extended structures under periodic boundary conditions. Therefore, the atomic pressure tensor is generally considered to be more useful, and it has been widely applied to study the mechanical properties of crystalline polymers,[180] lipid bilayers,[4,6,7] liposomes,[183] proteins,[8] metals and alloys.[184] In cases where molecules are small and behave like rigid bodies (such as water, methane, or ethylene), however, the molecular pressure tensor is more convenient. This is because the molecular formalism circumvents the evaluation of rigid constraints,[74,83,185] and thus potentially avoids the complexity of the anisotropic kinetic term in the atomic pressure tensor near interfaces.[186] Sega *et al.*[186] showed that, without considering such an anisotropic kinetic term for rigid molecules, the surface tension, calculated with only the configurational part of the atomic pressure tensor, tends to deviate from the true value by about 9 to 15% for water systems.

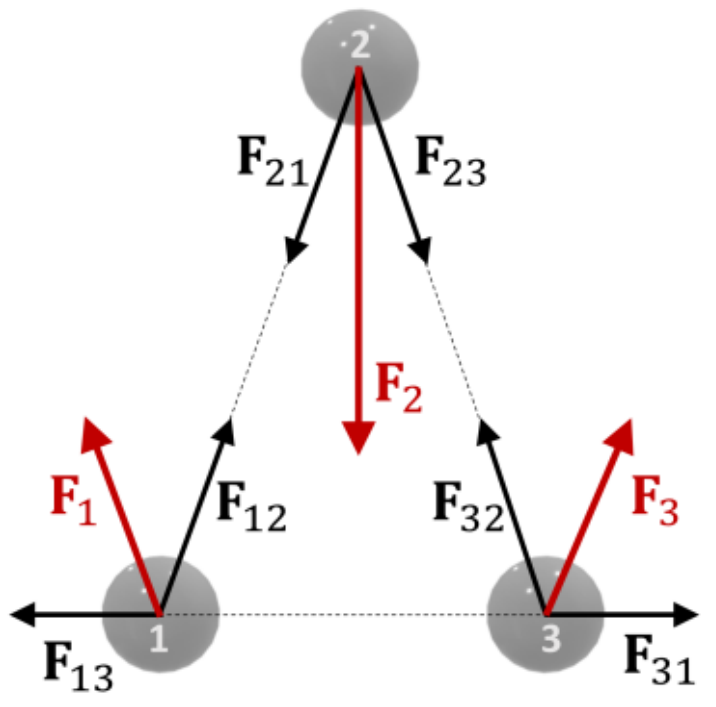

Figure 14. Central force (CF) decomposition for a three-body potential. In a CF decomposition,[78] the total force on an atom $\mathbf{F}_i$ is decomposed into pairwise central terms $\mathbf{F}_{ij}$; the (unknown) magnitude of $\mathbf{F}_{ij}$ can be obtained by solving a system of linear equations in Eq. (63). In a non-CF decomposition, such as the scheme proposed by Goetz and Lipowsky,[4] the pairwise term is simply $(\mathbf{F}_j - \mathbf{F}_i)/m$ with $m = 3$ for a three-body potential; this pairwise term is non-central, *i.e.*, pair-force is not parallel to the vector $\mathbf{r}_{ij}$.

The technical development for the atomic pressure/stress tensor has been centered on the decomposition of a many-body potential into pairwise components, so that the typical pairwise formalism based on Eq. (9) can be implemented. This force decomposition is not unique,[187] and a number of decomposition methods have been proposed.[4,78,176,183,188–191] A noticeable development is the so-called central force (CF) decomposition by Admal and Tadmor.[78] In a CF decomposition, the total force on atom $i$ due to a $m$-body potential $\mathcal{U}^{[m]}(\mathbf{r}^m)$ is written as a summation of pairwise central forces (*i.e.*, forces that are parallel to the vector $\mathbf{r}_{ij}$):

$$\begin{aligned} \mathbf{F}_i &= -\nabla_{\mathbf{r}_i} \mathcal{U}^{[m]}(\mathbf{r}^m) \\ &= \sum_{j(\neq i)} F_{ij} \frac{\mathbf{r}_{ij}}{r_{ij}} \end{aligned} \tag{63}$$

where $\mathbf{r}^m \equiv \mathbf{r}_1, \mathbf{r}_2, \ldots, \mathbf{r}_m$ is a collection of particle positions in a $m$-body cluster. Figure 14 illustrates the CF decomposition for a three-body potential. While some popular force decomposition methods violate the balance of linear[176,190,192] or angular[4] momentum, the CF decomposition yields a symmetric stress tensor by construction and satisfies the balance of both linear and angular momentum.[192] In practice, unknown parameters $F_{ij}$ are obtained by solving a system of linear equations given in Eq. (63). Here, the number of independent force equations is $3m - 6$ (as forces satisfy the conservation of linear and angular momentum), and the number of unknown pairwise central terms is $m(m-1)/2$. For $m = 3, 4$, solving this system of linear equations is a well-posed math problem. For potentials beyond four-body interactions, however, the number of unknown parameters is larger than the number of independent equations, and the CF decomposition becomes non-unique.[7] Torres-Sánchez *et al.*[191] developed a covariant central force (cCF) decomposition based on the Doyle-Ericksen relation of continuum mechanics, rather than on the statement of balance of linear momentum, as in the classical Irving-Kirkwood-Noll approach. The cCF decomposition is consistent with the CF approach for three- and four-body potentials but allows for many-body interactions of arbitrarily high order.[191,192] For pairwise potentials, all force decomposition schemes result in the same pressure or stress tensor. The non-unique scheme of the force decomposition is related to the non-uniqueness of the local pressure tensor due to arbitrary contour. One can argue that, using the thermodynamic route (Section 3.2), no force decomposition is needed, but the problem now becomes the ambiguity in assigning many-body potential energy into a local space.

The other technical challenge for both the atomic and molecular representations of the pressure tensor involve the consideration of the long-range Coulombic interactions. Unlike the short-range LJ potential, the Coulombic potential decays very slowly in space, and it is impossible to use a simple tail correction to account for the missing long-range part.[193] Compared to the direct Coulomb sum, which scales as $\mathcal{O}(N^2)$ (where $N$ is number of atoms or particles in the system), the Ewald summation method[194] is the standard method used in molecular simulations to efficiently handle the Coulombic interactions, scaling as $\mathcal{O}\left(N^{3/2}\right)$. Better efficiency can be achieved by modern, mesh-based Ewald methods,[195] which scale as $\mathcal{O}(NlogN)$. While the algorithm for computing the bulk (macroscopic) pressure tensor in the presence of the Coulombic interactions is well-established in the field,[179,196] it is still a challenge to handle the Coulombic or any long-range interactions in an efficient manner for the local pressure tensor. Previous studies have focused on finding a suitable scheme for assigning the local force that is compatible with the Ewald summation method. Since the $k$-space (Fourier space) part of the Coulombic energy in the Ewald method is the most computationally efficient in a non-pairwise form, the Harasima contour definition turns out to be a better choice than the IK contour. This can be understood by the delta function in the Harasima formulation (Eq. (16)); the delta function indicates that the configurational part only contributes to the tangential pressure at planes where molecules are present. This feature allows the per-atom form of the $k$-space energy term to be naturally incorporated into the Harasima formulation. Compatibility of the Ewald summation method with the Harasima contour has been developed for the local tangential pressure across a planar interface[6,185,197] and for the local axial pressure in a cylindrical geometry[74]. Nevertheless, it is still possible to use the IK contour with the Ewald method. Hatch and Debenedetti[8] successfully captured the full Coulombic energy using the IK contour definition by writing the Ewald sum in an explicit pairwise form. However, as expected, a pairwise form of the Ewald sum is computationally expensive. A computationally amenable alternative using the IK contour is considering the Coulombic potential up to a certain cutoff radius,[7,183] but such a treatment cannot guarantee a consistent pressure profile because the Coulombic potential was treated differently (with Ewald-based method) in the molecular simulations and (with simple cutoff) in the pressure calculation. It is possible to replace the bare Coulombic potential with a damped, shifted-potential[198] or a shifted-force[199] one, which is generally called the Wolf potential. The Wolf potential can reproduce the energetics and dynamics of various systems to an acceptable accuracy compared to the (exact) Ewald method. However, the error that might be introduced to the pressure tensor by this (approximate) Wolf potential is still unclear. We note that the preference of the Harasima contour in the treatment of the Coulombic interactions does not imply that the Harasima contour is more correct than the IK contour, but the Harasima/Ewald method clearly has advantages in its computational efficiency.

Future breakthrough in this regard would rely on a better understanding of the physical nature of the microscopic pressure tensor in the presence of multi-body and Coulombic interactions. The possibility to define a unique CG pressure tensor (Section 5.1) will benefit the calculation of Coulombic contributions to the pressure tensor, as we are free to choose the most convenient contour (e.g., Harasima contour) without worrying the physical and numerical arbitrariness of the final results. Due to the convenience in calculating the short-range interaction, it is also worthwhile to systematically evaluate the uncertainty that a spherically truncated or damped Coulombic potential will introduce to the pressure in general, in comparison to the exact treatment using direct Coulomb sum (without potential cutoff), or more efficient Ewald-based methods.

## 5.3 Software and computational tools

Currently, the LAMMPS molecular dynamics software provides native functionalities for calculating the local and global stress/pressure tensor on-the-fly, and these *compute* commands are summarized in Table 1. In contrast, the NAMD software only provides a basic function to compute the pressure tensor profile along the *z*-axis for systems having a planar geometry.[200] This specific implementation adopts the Harasima contour definition, which allows an efficient computation of the Coulombic contribution to the local pressure tensor based on the Ewald method (see Section 5.2 for details).[6] The limited pressure or stress tensor functionality in major molecular simulation software has motivated the development of dedicated analysis tools. Nakamura *et al.*[72,201] prepared a patch file for the LAMMPS software that enables the calculation of the local pressure tensor in Cartesian and spherical coordinates. Vanegas *et al.*[191,202] developed a computational tool "GROMACS-LS" for the GROMACS software to calculate the local stress in molecular systems. Admal *et al.*[78,203] developed a post-analysis program "MDStressLab" that takes the input data (particle coordinates, velocities, species etc.) in a general format, so it is compatible with different simulation software provided that the scripts for converting software-specific file format are available. All of these dedicated analysis tools are capable of calculating the 3D pressure or stress tensor field (as a volume average value, Hardy definition) in systems having arbitrary geometries. They differ in the availability of pressure/stress tensor definitions, force decomposition schemes for many-body potentials, and supported interaction potentials. Interested readers are referred to the program's user guide for details. Other pressure tensor codes that were designed for specific coordinate systems and intermolecular potentials are also available.[204–206]

At the moment, all available computational tools are more or less limited in definitions of the local pressure tensor, in applicability to certain system geometries, and in availability of intermolecular potentials. Therefore, there is a strong motivation to develop a general-purpose software in the future with a well-documented user guide. As users may have personal preference for molecular simulation packages, it is imperative to develop compatibility of this pressure analysis software with different simulation packages. This could be realized through support for a range of input formats (or at least, scripts for converting file formats), or a generalized cross-language functional interface that can be called from any of the codes during the force calculation to tally the pressure. Calculating the local pressure tensor is computationally expensive, therefore parallelization of the computation using the high-performance central/graphics processing unit (CPU/GPU) would be essential in the future.

The development of a standard computational package for the microscopic pressure tensor will also standardize the measurement of pressure and enable the generation of a database of high-fidelity pressure/stress data for systems of both engineering and scientific interest. The availability of a large amount of data organized in a curated database will help advance the process optimization and materials design using machine learning and data science.

Table 1. Pressure or stress tensor functionalities available in the LAMMPS software. Note that the stress tensor is defined as the negative of the pressure tensor. More detailed explanations and restrictions of these commands are available in the LAMMPS manual (*https://docs.lammps.org/Manual.html*, accessed on May 16, 2022, LAMMPS version 4 May 2022)

| LAMMPS command | Explanation |
|---|---|
| compute stress/atom | Computation of per-atom stress tensor. A virial contribution produced by a *m*-body potential is equally assigned to each atom in the set, *e.g.*, 1/4 of the dihedral virial to each of the 4 atoms. We note that this function is commonly used for visualization purpose; caution should be exercised when interpreting this per-atom value in a local fashion which has some similarities to the IK1 approximation (Eq. (7)). This command works for long-range[179] and many-body interactions[207]. |
| compute pressure | Computation of a scalar pressure and a global pressure tensor of the entire system.[207] |
| compute pressure/uef | Computation of the pressure tensor in the reference frame of the applied flow field. |
| compute stress/mop[a] | Computation of local stress tensor using the method of planes (Eq. (54)). Specifically, it computes 3 components in directions $\alpha x$, $\alpha y$ and $\alpha z$, where $\alpha$ is the direction normal to the plane.[61] The profile of the stress can be computed with "compute stress/mop/profile" command. |
| compute stress/cartesian[a,b] | Computation of coarse-grained profiles of the diagonal components of the local stress tensor in Cartesian coordinates. The output stress tensor is averaged over a small local volume (Eq. (52) with a slab-like local volume).[175] |
| compute stress/spherical[a,b] | Computation of profiles of the diagonal components of the local stress tensor in spherical coordinates (Eq. (52) with a spherical-shell local volume).[175] |
| compute stress/cylinder[a,b] | Computation of profiles of the diagonal components of the local pressure tensor in cylindrical coordinates (Eqs. (28), (30), and (32)).[21] This command does not consider periodic boundary conditions, so the system should be large enough to ensure the boundary effect is negligible. |

[a] The command works only for short-range pair interactions; *i.e.*, if any bond, angle, dihedral, *etc.* contributions and *k*-space contributions (in Ewald summation method) for the long-range Coulombic interactions are present in the system, the results will be incorrect.
[b] The calculation is based on the IK contour definition.

## 5.4 Experimental measurements of microscopic pressure tensor

At the moment, it is still a challenge to directly measure or estimate the pressure or stress tensor at the nanoscale from experiments, and nearly all microscopic pressure or stress tensor results reported in the literature so far are theoretical values. The difficulties come from the limitation in experimental techniques to approach the molecular level stress, and from the non-uniqueness in the definition of the microscopic pressure tensor. Advances in either side will significantly benefit the other.

Gubbins and co-authors have demonstrated that the pressure tensor in a molecular scale system could be estimated from experimental input in simple equilibrium systems, such as for fluids adsorbed in a carbon slit-shaped pore. The normal pressure component, $P_N$, in such a system is a constant, and Śliwińska-Bartkowiak and co-authors[208,209] demonstrated that it is possible to estimate this pressure by measuring the resulting changes in the interplanar distance of the activated carbon fibers, using X-ray diffraction. The in-pore normal pressure can then be estimated using Young's equation, provided that the transverse compressive modulus is known. Figure 15 shows that an agreement has been achieved between the simulated normal pressure and the experimental estimations, within the (rather large) uncertainties of the latter, for $CCl_4$ and $H_2O$ adsorbed in carbon slit pores.

For the tangential pressure, $P_T$, the experimental estimation is more challenging because the local tangential pressure is non-unique, and the force does not act directly on the adsorbent material, but in a direction parallel to the wall. Thus, a non-invasive method is needed. Molecular simulation and experimental results show that for adsorbates that wet the pore walls the adsorbed layers of molecules very near to the wall are quasi-two-dimensional; this being particularly pronounced for the contact layers next to the pore walls. This observation enabled the development of a "2D route" to the effective tangential pressure inside a single adsorbed layer.[210,211] In this 2D route, the behavior of a single adsorbed layer near the surface is related to that of a strictly 2D reference film by projecting the center of mass positions of the molecules in the layer onto the surface plane. The 2D pressure $P_{2D}$ (in units of force per unit length) in the reference film is then mapped back to the 3D pressure (in units of force per unit area) by being divided by an effective length scale $l_{eff}$ in the direction normal to the surface. The effective tangential pressure estimated by the 2D route is:[211]

$$\overset{\text{2D}}{P_T} \equiv \frac{P_{2D}(T, \rho_{2D})}{l_{eff}} \qquad (64)$$

where $T$ is temperature and $\rho_{2D}$ is the 2D density of the adsorbed film. The 2D pressure $P_{2D}$ is a function of $T$ and $\rho_{2D}$, and the relation can be established by a 2D equation of state.[210] Although $P_{2D}$ is well defined, $\overset{\text{2D}}{P_T}$ is not unique due to the arbitrary choice of $l_{eff}$. To decide on a sensible choice of $l_{eff}$, it is instructive to rewrite Eq. (64)(A11) as a spatial (volume) average,[211] *i.e.*, $\overset{\text{2D}}{P_T} \approx \int_{l_{eff}} P_T(z)\mathrm{d}z \,/ l_{eff}$. Based on the results in Figure 13, it is motivating to choose $l_{eff}$ to be the characteristic length $\Delta z_k$ that is comparable to the thickness of an adsorbed layer, so that the spatial average appears to be unique and has a clear physical meaning. In this 2D route, experimental input parameters are $T$, $\rho_{2D}$, and $l_{eff}$. The 2D density $\rho_{2D}$ can be estimated from molecular simulations or adsorption theories, or obtained from particular experiments such as small-angle neutron diffraction. The effective thickness of the layer $l_{eff}$ can be estimated from molecular simulations and theories, or from optical sensing experiments[212]. Limitations of the 2D route are: 1). It neglects the interactions between the layer of interest and its neighboring layers. 2) It may fail when the adsorbed layer deviates from a quasi-2D structure, for example, in weakly wetting systems or for materials with a rough surface.

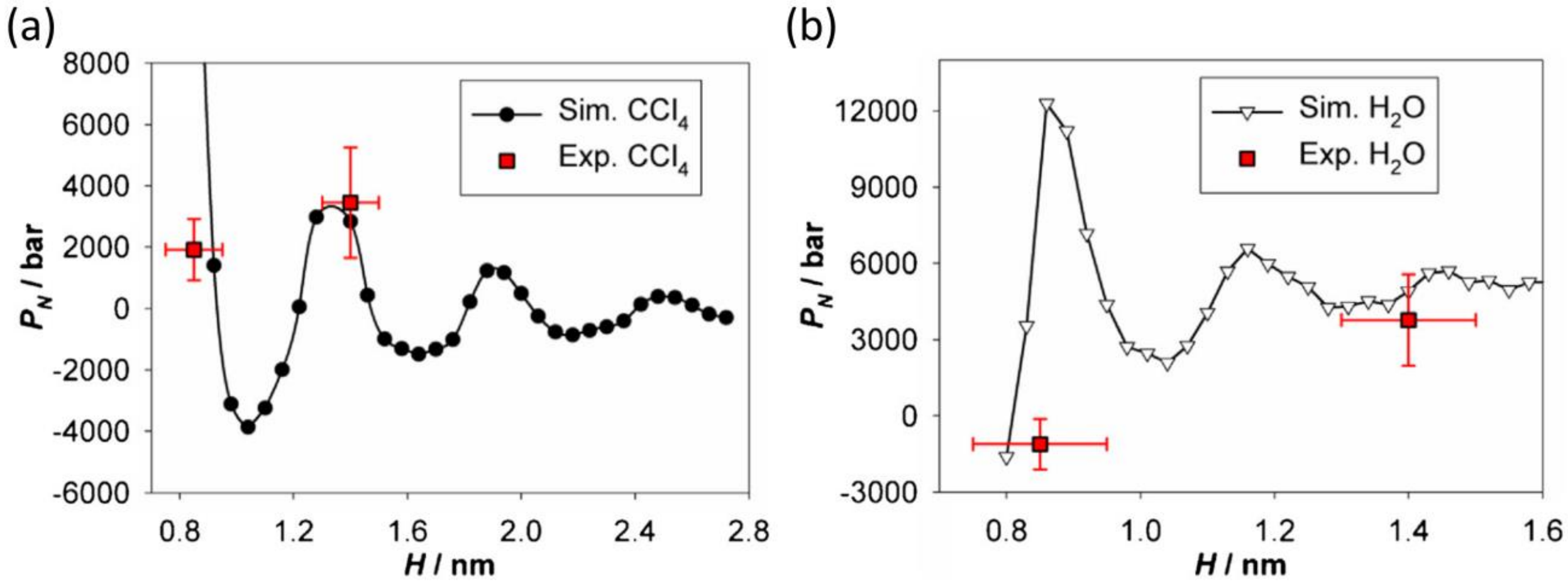

Figure 15. Molecular simulation and experimental results for the normal pressure of (a) $CCl_4$ and (b) $H_2O$ adsorbed in slit carbon pores of different pore width $H$, at 300 K and 1 bar bulk pressure. The experimental data were estimated using Young's equation based on the change of interplanar distance of the activated carbon fibers. Adapted with permission from Ref. [209], Copyright 2013 Elsevier.

Advanced experimental techniques have been developed to estimate the scalar pressure at very small scales. These techniques could serve as a foundation for future routes to approach the microscopic pressure in its correct tensor form:

- **Direct methods**. The pressure can be directly estimated by measuring the mechanical response of the system to a physical probe, such as an atomic force microscopy (AFM) tip.[213–215] This method has been commonly adopted to estimate the scalar pressure inside a nanobubble.[213,215] For example, the interaction force between the nanobubble and the AFM tip can be dynamically captured based on the AFM cantilever's instantaneous deflection. The internal pressure of the nanobubble can be obtained by fitting the recorded force-displacement curve to a theory that separates the internal pressure of the confined fluids from the elastic deformation of the solid materials.[213] Another possible way is to relate the pressure to the elastic properties of the system, such as the elastic modulus, $K_T = \beta_T^{-1}$, where the isothermal compressibility $\beta_T$ is defined as

$$\beta_T = -\frac{1}{V}\left(\frac{\partial V}{\partial P}\right)_T \tag{65}$$

$\beta_T$ of the confined fluids can be measured by ultrasonic experiments.[216,217] However, the measured elastic property is commonly interpreted as a macroscopic (scalar) quantity averaged over all directions and the system volume, instead of as a tensor.

- **Indirect methods**. The scalar microscopic pressure can be sensed by molecular probes that are (mechanically, optically, thermally etc.) susceptible to the pressure change of the environment. Vasu *et al.*[28] studied the effect of confinement between two graphene layers on molecules that are susceptible to deformation under pressure. By comparing Raman spectra for the confined molecules with those for the same molecules in the bulk phase, they estimated the "van der Waals" (scalar) pressure in the confined phase to be 1-1.5 GPa. Another relevant example is using molecular rotors as sensors to measure the local viscosity of a fluid under extreme confinement conditions.[218] The probe molecular rotor changes its fluorescent properties due to the pressure and shear of the surrounding fluids. This information could allow an estimate of the pressure tensor in a fluid away from equilibrium at very small scales.

An immediate challenge is the lack of rigorous relations that connect the experimental measurements to the microscopic pressure tensor in statistical mechanics. Future studies should focus on establishing these fundamental relations to bridge the gap between the experiments and theories.

# 6. Concluding Remarks

In this perspective, we have reviewed several routes to calculate the microscopic pressure tensor (equivalent to the negative stress tensor), in both equilibrium and non-equilibrium systems. These formulae can be generally divided into two types, depending on whether they were derived via mechanical or thermodynamic routes. The mechanical (or "virial") route follows the mechanical concept of "the force acting across a surface", and it can be used in both equilibrium and non-equilibrium systems. By contrast, the thermodynamic route uses the thermodynamic definition of the pressure, which is the negative of the change of the Helmholtz free energy with respect to volume. The thermodynamic route can only be used in equilibrium fluid systems where no shearing is present, but it is arguably preferable to the mechanical route for systems interacting with complex (multi-body) potentials.

We have categorized available pressure equations into different forms, based on where and how the microscopic pressure tensor is measured. These include macroscopic (bulk), pointwise, volume and surface forms. We then attempt to show the underlying connection between the different forms, highlighting the inherent assumptions with the limitations of each of these choices. The equations and connections between different forms are summarized in Figure 16 for the configurational part of the pressure tensor and in Figure 17 for the corresponding kinetic part.

We have also pointed out four aspects that currently face challenges and need further investigations. In brief, they are:

- **Historical controversies over the definition of the microscopic pressure tensor.** Controversies are centered on the non-uniqueness of the microscopic pressure tensor at a point in space, resulting from the fact that the forces between molecules do not act at a unique point in space; this difficulty manifests itself in the equations as the arbitrary contour involved in the calculations. However, coarse-graining over a relatively small spatial region of space may result in a well-defined pressure tensor, as has been shown recently for a simple system.[35] A breakthrough in this regard may open the door to a thermodynamically and mechanically consistent picture of the nanoscale systems.

- **Difficulties with many-body and long-range potentials.** This technical difficulty is outstanding for complex systems, such as those in biology. Future research is suggested to focus on better understanding of the physical nature of the pressure tensor in the presence of many-body interactions, and on the development of convenient, accurate, and efficient algorithms to account for long-range interactions in the pressure tensor.

- **Inadequate software and computational tools for the calculation of the local pressure/stress tensor.** To our knowledge, no general-purpose software is available. Such software and computational tools are essential to avoid confusion, and to overcome the knowledge barriers for non-expert users, and are needed to accelerate the process optimization and materials design using machine learning and data science. The ease of calculating the virial pressure results in its use when running exiting software packages, often in cases where it is invalid (e.g. confined flows).

- **Lack of experimental methods to measure the pressure tensor at the nanoscale.** Advances in this regard require a combined effort from both the experimental and theoretical/computational communities. A breakthrough in measuring the microscopic pressure tensor will enable the determination of a wide range of thermodynamic and transport properties at the nanoscale in the correct tensor form.

The microscopic pressure tensor is a pivotal property for a large range of disciplines and technologies, including fluid dynamics, solid mechanics, biophysics, thermodynamics, nucleation and crystallization, chemical manufacturing, and chemical separations. We invite both experimentalists and theorists to contribute to this field to enable a thorough and consistent understanding of tensorial features of inhomogeneous systems.

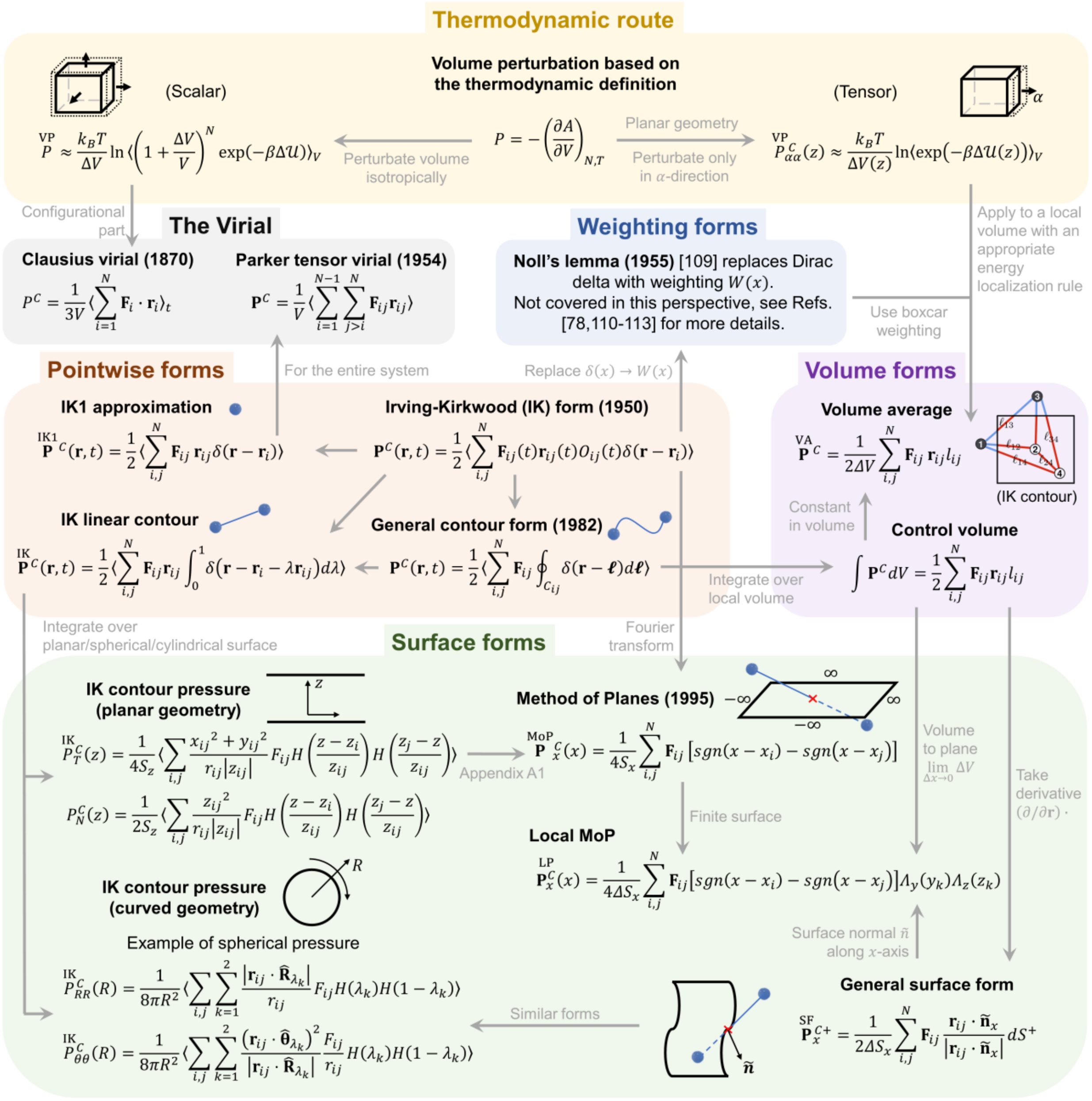

Figure 16. A schematic knowledge flowchart relating different configurational pressure forms discussed in this perspective. See Figure 17 for a similar flowchart for the kinetic pressure. Any forms without angular brackets can be applied away from equilibrium.

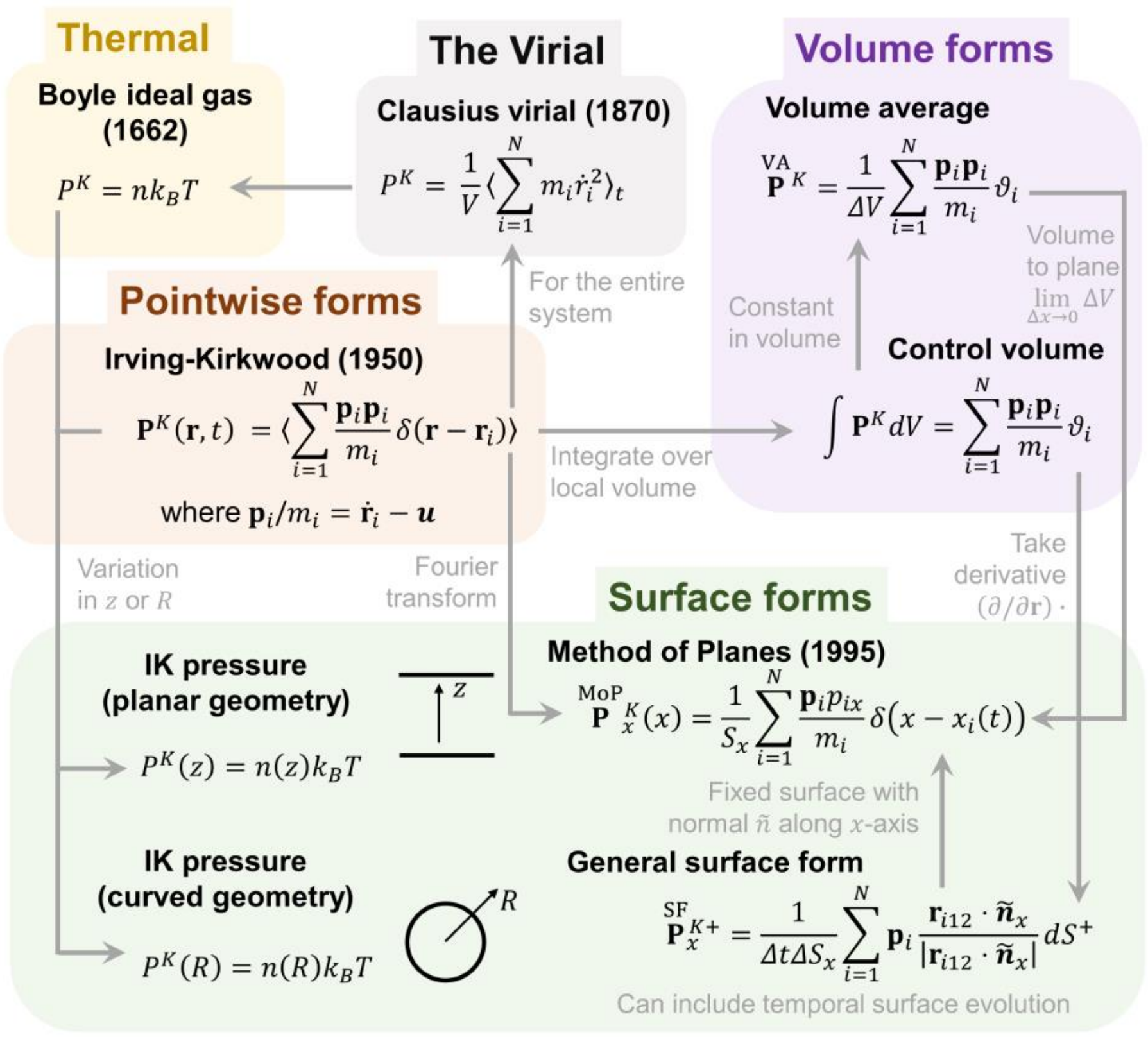

Figure 17. A schematic knowledge flowchart relating different kinetic pressure forms discussed in this perspective. See Figure 16 for a similar flowchart for the configurational pressure.

# Appendix

## A1. Linking the IK-contour pressure to the MoP form

In this appendix, we derive the mathematical relationship between the configurational part of the IK-contour pressure (Eqs. (13) and (15)) and the MoP pressure (Eq. (54)). This is significant as the MoP pressure is derived in Fourier space and valid away from equilibrium, while the IK-contour pressure is obtained by integrating (averaging) over the $x$ and $y$-directions and generally used only in an equilibrium system. Considering only the configurational part of Eq. (13) and splitting Eq. (15) into two the different tangents, with averaging notation $\langle ... \rangle$ omitted for simplicity, we have

$$\overset{IK}{P}{}^C_N(z) = \frac{1}{2S_z}\sum_{i,j}^{N}\frac{z_{ij}{}^2}{r_{ij}}\frac{1}{|z_{ij}|}F_{ij}H\left(\frac{z - z_i}{z_{ij}}\right)H\left(\frac{z_j - z}{z_{ij}}\right) \quad \text{(A1)}$$

$$\overset{\text{IK}}{P}{}^{C}_{Tx}(z) = \frac{1}{2S_z}\sum_{i,j}^{N}\frac{{x_{ij}}^2}{r_{ij}}\frac{1}{|z_{ij}|}F_{ij}H\left(\frac{z - z_i}{z_{ij}}\right)H\left(\frac{z_j - z}{z_{ij}}\right) \tag{A2}$$

$$\overset{\text{IK}}{P}{}^{C}_{Ty}(z) = \frac{1}{2S_z}\sum_{i,j}^{N}\frac{{y_{ij}}^2}{r_{ij}}\frac{1}{|z_{ij}|}F_{ij}H\left(\frac{z - z_i}{z_{ij}}\right)H\left(\frac{z_j - z}{z_{ij}}\right) \tag{A3}$$

For comparison, we rewrite the MoP pressure in Eq. (54) for three components on the $z$-surface (the surface that is normal to the $z$-direction); the configurational term gives,

$$\overset{\text{MoP}}{\mathbf{P}}{}^{C}_{z}(z) = \frac{1}{4S_z}\left[\sum_{i,j}^{N}\mathbf{F}_{ij}\left[sgn(z - z_i) - sgn(z - z_j)\right]\right] \tag{A4}$$

Noting that the scalar force in Eqs. (A1)-(A3) can be related to its vector form by $\mathbf{F}_{ij} = F_{ij}\mathbf{r}_{ij}/r_{ij}$ and the projection of the pair force in the $\alpha$-direction ($\alpha = x, y, z$) is $F_{ij\alpha} = F_{ij}\alpha_{ij}/r_{ij}$, we write Eqs. (A1)-(A3) in a unified vector form as,

$$\overset{\text{IK}}{\mathbf{P}}{}^{C}_{z}(z) = \frac{1}{2S_z}\sum_{i,j}^{N}\mathbf{F}_{ij}\circ\mathbf{r}_{ij}\frac{1}{|z_{ij}|}H\left(\frac{z - z_i}{z_{ij}}\right)H\left(\frac{z_j - z}{z_{ij}}\right) \tag{A5}$$

where $\mathbf{F}_{ij}\circ\mathbf{r}_{ij}$ denotes the element-wise product of force vector $\mathbf{F}_{ij}$ and distance vector $\mathbf{r}_{ij}$.The Heaviside function expressed as a product gives identical behavior to the difference of two Heaviside functions:

$$H\left(\frac{z - z_i}{z_{ij}}\right)H\left(\frac{z_j - z}{z_{ij}}\right) = H\left(\frac{z - z_i}{z_{ij}}\right) - H\left(\frac{z - z_j}{z_{ij}}\right) \tag{A6}$$

Using the property $H(ax) = 1/2\,(sgn(a)sgn(x) + 1)$ to transform Eq. (A6) to signum functions, we get

$$H\left(\frac{z-z_i}{z_{ij}}\right)-H\left(\frac{z-z_j}{z_{ij}}\right) = \frac{1}{2}\left[sgn\left(\frac{1}{z_{ij}}\right)sgn(z-z_i)+1\right] - \frac{1}{2}\left[sgn\left(\frac{1}{z_{ij}}\right)sgn(z-z_j)+1\right] = \frac{1}{2}sgn\left(\frac{1}{z_{ij}}\right)\left[sgn(z-z_i)-sgn(z-z_j)\right] \tag{A7}$$

Using the definition of the signum function $sgn(z_{ij}) = z_{ij}/|z_{ij}|$, the product is $sgn\left(1/z_{ij}\right)sgn\left(z_{ij}\right) = 1$, so Eq. (A1) can be rewritten as

$$\overset{\text{IK}}{P}{}^{C}_{N} = \overset{\text{IK}}{P}{}^{C}_{zz} = \frac{1}{2S_z}\sum_{i,j}^{N}\frac{z_{ij}}{|z_{ij}|}F_{ijz}\frac{1}{2}\text{sgn}\left(\frac{1}{z_{ij}}\right)\left[sgn(z-z_i)-sgn(z-z_j)\right] = \frac{1}{4S_z}\sum_{i,j}^{N}F_{ijz}\left[sgn(z-z_i)-sgn(z-z_j)\right] = \overset{\text{MoP}}{P}{}^{C}_{zz} \tag{A8}$$

Here we obtain the equivalence between the normal MoP pressure on a $z$-surface and the corresponding $zz$-component in the IK-contour pressure tensor (and note $zz$-component is independent of the contour definition).

The tangential components, however, highlight a fundamental difference between the IK-contour pressure and the MoP pressure. For example, considering the $yy$-component of Eq. (A3) and rewriting it in the same way as Eq. (A8), it is trivial to show,

$$\overset{\text{IK}}{P}{}^{C}_{Ty} = \overset{\text{IK}}{P}{}^{C}_{yy} = \frac{1}{4S_z}\sum_{i,j}^{N}\frac{y_{ij}}{z_{ij}}F_{ijy}\left[sgn(z-z_i)-sgn(z-z_j)\right] \tag{A9}$$

where we have an extra factor of $y_{ij}/z_{ij}$ when compared to the normal MoP pressure on a $y$-surface (similar to Eq. (56)). The IK-contour pressure therefore uses extra molecular information by taking the $y_{ij}$ distance to calculate the direct $y$-pressure on a $z$-plane. This is, however, a departure from the Cauchy definition of pressure as shown in Figure 4a, which requires the three different normal pressures to be defined on orthogonal planes. Other form of the pressure, for example the VA pressure, uses this same quantity $F_{ijy}y_{ij}$ to get $P_{yy}$, but in the VA form, this is weighted by the fraction of interaction $l_{ij}$ inside the volume. In contrast, the IK-contour form accumulates interactions crossing a $z$-plane and weights these by the length of interaction in the surface normal direction $z_{ij}$ as shown in Figure A1. We also note that averaging the IK-contour pressure over a volume should lead to the corresponding VA pressure form.

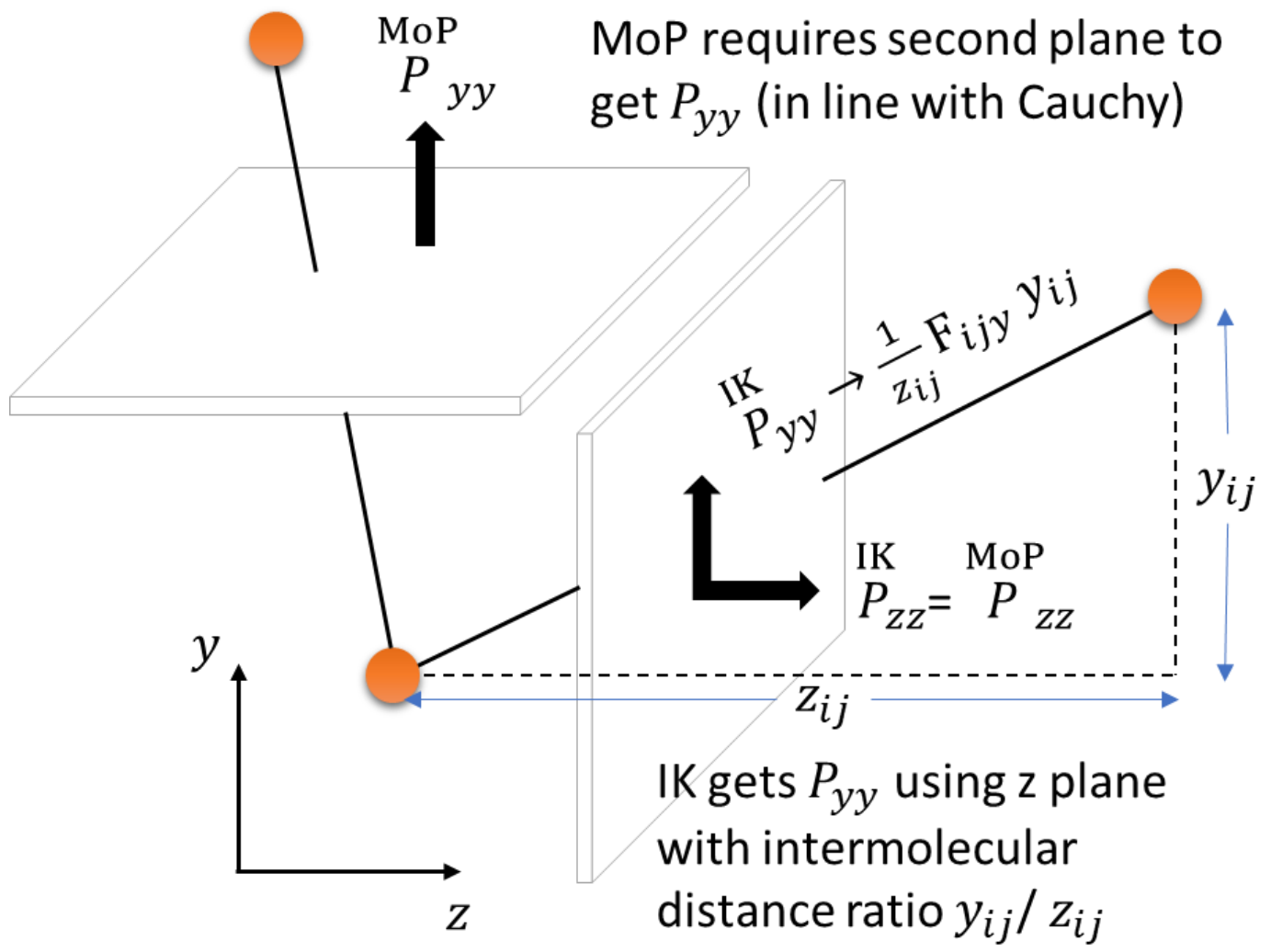

Figure A1. A schematic showing $P_{yy}$ and $P_{zz}$ on a single $z$-normal plane from the IK-contour method, while the MoP pressure defines a $y$-normal plane in order to get $P_{yy}$.

## A2. The Noll form of Pressure

The Noll reformulation replaces the Dirac delta function $\langle\delta(\mathbf{r}-\mathbf{r}_i)\rangle \rightarrow \langle f \,|\, \mathbf{r}_i = \mathbf{r}\rangle$. This notation denotes an integral of probability density function $f$ over all phase space except $\mathbf{r}_i$ which is then replaced by $\mathbf{r}$. Conditions are then placed on $f$, including a similar condition to the phase space bounded assumption of Irving and Kirkwood. The form of configurational pressure uses Noll's Lemma to reformulate the IK operator in his notation is,

$$\overset{\text{NOLL}}{\mathbf{P}}{}^{C}(\mathbf{r},t) = -\frac{1}{2}\sum_{i,j}^{N}\int_{\boldsymbol{z}} \boldsymbol{z}\int_0^1 \langle \mathbf{F}_{ij}\, f \,|\, \mathbf{r}_i = \mathbf{r}+\lambda\boldsymbol{z}, \mathbf{r}_j = \mathbf{r}-(1-\lambda)\boldsymbol{z}\rangle d\lambda d\boldsymbol{z} \qquad \text{(A10)}$$

This states that the stress tensor at point $\mathbf{r}$ is a superposition of expectation forces from all possible bonds that might pass point $\mathbf{r}$, while the integral over $\lambda$ "slides" from $\mathbf{r}_i$ to $\mathbf{r}_j$ along a vector $\boldsymbol{z}$ between the molecules (See Ref. [78] for a sketch). This form is potentially more general than the pair potential which is assumed by Eq. (8) where $\boldsymbol{z} = \boldsymbol{\ell}$ can be a general contour.[78] For the case of pairwise interactions, $\mathbf{z} = \mathbf{r}_{ij} = \mathbf{r}_j - \mathbf{r}_i$, we can rearrange both equalities in the brackets to show they are the same condition $\mathbf{r}_j - (1-\lambda)\mathbf{r}_{ij} = \mathbf{r}_i + \lambda\mathbf{r}_{ij}$,

$$\overset{\text{NOLL}}{\mathbf{P}}{}^{C}(\mathbf{r},t) = \frac{1}{2}\sum_{i,j}^{N}\int_{\boldsymbol{r}_{ij}} \mathbf{r}_{ij}\int_0^1 \langle \mathbf{F}_{ij}\, f \,|\, \mathbf{r}_i + \lambda\mathbf{r}_{ij} = \mathbf{r}\rangle d\lambda d\mathbf{r}_{ij} \qquad \text{(A11)}$$

which is similar to the form of line integral used throughout this work. Note in general that we depart from the approach of Noll, as the equations we consider are derived without the use of an ensemble average[41,110,112] as discussed in Section 4.

## Acknowledgment

KS thanks Dr. Yun Long for helpful discussions about the thermodynamic route for curved interfaces. KS also thanks Dr. Harold ('Wick') Hatch for pointing out important references on the canonical transformation. ES acknowledges funding from the EPSRC project EP/S019545/1 and is grateful to the UK Materials and Molecular Modelling Hub for computational resources, which is partially funded by EPSRC (EP/P020194/1 and EP/T022213/1). KEG thanks the U.S. National Science Foundation for partial support under grant no. CBET-1603851. EES thanks the U.S. National Science Foundation for support under grant no. CBET-1855465.

## Conflict of Interest

The authors have no conflicts to disclose.

## Data Availability

The data that support the findings of this study are available from the corresponding author upon reasonable request.